\documentclass[12pt]{article}

\usepackage{graphics}
\usepackage{amssymb}

\newcommand{\bea}{\begin{eqnarray}}
\newcommand{\eea}{\end{eqnarray}}

\newcommand{\e}{\mbox{e}}

\newcommand{\ra}{\rangle}
\newcommand{\la}{\langle}

\newcommand{\mi}{\!-\!}
\newcommand{\equ}{\!=\!}
\newcommand{\pl}{\!+\!}

\newcommand{\R}{{\rm I\!R}}
\def\void{}
\def\labelmark{}

\newenvironment{formula}[1]{\def\labelname{#1}
\ifx\void\labelname\def\junk{\begin{displaymath}}
\else\def\junk{\begin{equation}\label{\labelname}}\fi\junk}%
{\ifx\void\labelname\def\junk{\end{displaymath}}
\else\def\junk{\end{equation}}\fi\junk\labelmark\def\labelname{}}

{\ifx\void\labelname\def\junk{\end{array}\end{displaymath}}
\else\def\junk{\end{array}\right.\end{equation}}
\fi\junk\labelmark\def\labelname{}\def\junk{}
\def\arraystretch{1}}

\newcommand{\beq}{\begin{formula}}
\newcommand{\eeq}{\end{formula}}
\newcommand{\beqv}{\begin{formula}{}}

\begin{document}

\hfill AEI-2001-049
\vspace{.4cm}

\hfill  24 May 2001

\begin{center}
\vspace{24pt}
{\Large\bf Dynamically Triangulating Lorentzian Quantum Gravity}

\vspace{30pt}

{\sl J. Ambj\o rn}$\,^{a}$,
{\sl J. Jurkiewicz}$\,^{b}$ and
{\sl R. Loll}$\,^{c}$\footnote{address from Sept `01: 
Institute for Theoretical Physics, Utrecht University, Minnaertgebouw,
Leuvenlaan 4, NL-3584 CE Utrecht}

\vspace{18pt}

$^a$~The Niels Bohr Institute, \\
Blegdamsvej 17, DK-2100 Copenhagen \O , Denmark\\
{\it email: ambjorn@nbi.dk}

\vspace{10pt}
$^b$~Institute of Physics,\\
Jagellonian University,\\
Reymonta 4, PL 30-059 Krakow, Poland\\
{\it email: jurkiewi@thrisc.if.uj.edu.pl}

\vspace{10pt}
$^c$~Albert-Einstein-Institut,\\
Max-Planck-Institut f\"{u}r Gravitationsphysik,\\
Am M\"uhlenberg 1, D-14476 Golm, Germany\\
{\it email: loll@aei-potsdam.mpg.de}

\vspace{24pt}

\end{center}


\begin{center}
{\bf Abstract}
\end{center}

Fruitful ideas on how to quantize gravity are few and far
between. In this paper, we give a complete description of a
recently introduced non-perturbative gravitational path integral
whose continuum limit has already been investigated extensively 
in $d<4$, with promising results. It is based on a simplicial 
regulari\-zation of Lorentzian space-times and, most importantly, 
possesses a well-defined, non-perturbative Wick rotation. 
We present a detailed analysis of the geometric and mathematical
properties of the discretized model in $d=3,4$. This includes a 
derivation of Lorentzian simplicial manifold constraints, the
gravitational actions and their Wick rotation. We define a transfer
matrix for the system and show that it leads to a well-defined
self-adjoint Hamiltonian. In view of numerical simu\-la\-tions,
we also suggest sets of Lorentzian Monte Carlo moves. We demonstrate
that certain pathological phases found previously in Euclidean
models of dynamical triangulations cannot be realized in the
Lorentzian case.

\vspace{12pt}
\noindent


\newpage

\section{Introduction}\label{intro}

Despite an ever-increasing arsenal of sophisticated mathematical 
machinery, the non-perturbative quantization of gravity remains 
an elusive goal to theorists. Although seemingly 
negligible in most physical situations, quantum-gravitational phenomena
may well provide a key to a more profound understanding of nature. 
Unfortunately, the entanglement of technical problems
with more fundamental issues concerning the structure of a
theory of quantum gravity often makes it difficult to pinpoint why 
any particular quantization program has not been successful.

The failure of perturbative methods to define a fundamental 
theory that includes gravity has led to
alternative, non-perturbative approaches which seek to describe
the quantum dynamics of gravity on the mother of all spaces, the
``space of geometries''. By a ``geometry'' we mean a space-time
with Lorentzian metric properties. Classically, such space-times 
usually come as smooth manifolds $M$ equipped with a metric tensor
field $g_{\mu\nu}(x)$. Since any two such metrics are 
physically equivalent if they can be mapped onto each other 
by a diffeomorphism of $M$, the physical degrees of freedom 
are precisely the ``geometries'', namely, the equivalence classes of 
metrics with respect to the action of Diff($M$).

In what follows, we will describe a path-integral approach to quantum 
gravity that works much more directly with the ``geometries''
themselves, without using coordinates\footnote{There is
no residual gauge invariance since the state sum is taken over
inequivalent discretized geometries. In this sense, the formalism is 
manifestly diffeomorphism-invariant. In Euclidean 2d quantum gravity,
there is ample evidence that this procedure is equivalent to a gauge-fixed 
continuum formulation, see \cite{ginsmoo,david} for reviews.}. 
It is related in spirit 
to other constructions based on simplicial ``Regge'' geometries 
\cite{reggerev}, in particular, the method of dynamical triangulations
\cite{dyntri,iceland} 
(see \cite{livrev} for a comparison and critical appraisal
of various discrete approaches to 4d quantum gravity). 
Its main aim is to construct an interacting quantum theory
of gravity in four dimensions as the continuum limit of a well-defined 
statistical mechanics model of regularized gravitational 
field configurations. The formal continuum path integral for gravity
is represented by a discrete sum over inequivalent triangulations $T$,
\begin{equation}
Z=\int_{\frac{\rm Metrics}{\rm Diff}}{\cal D}g\ {\rm e}^{iS[g]}\; 
\hookrightarrow\; 
Z= \sum_{T}m(T)\ {\rm e}^{iS(T)},
\label{pathint}
\end{equation}
where $m(T)$ is a measure on the space of discrete geometries.
Unlike previous approaches, we use a space of piecewise linear 
{\it Lorentzian} space-times as our starting point. 
That this procedure is in general inequivalent to path integrals over 
bona fide Euclidean geometries has already been demonstrated 
in two dimensions \cite{al,alnr}. 

The reason why previous discrete path-integral approaches have been
formulated for {\it Euclidean} gravity\footnote{Euclidean gravity 
is the theory with (minus the)
action (\ref{contact}) below, where the Lorentz metric $g_{\mu\nu}$ 
has been substituted by a Riemannian metric $g_{\mu\nu}^{\rm eucl}$ 
with positive-definite signature.} 
is often a technical one, 
rather than the conviction that Euclidean space-times are more
fundamental physical quantities than Lorentzian ones
(after all, a phenomenon like black holes is intimately related
to the causal, Lorentzian structure). The corresponding 
real weight factors e$^{-S^{\rm eucl}}$ are then used 
in Monte Carlo simulations, and also usually necessary (though not
sufficient) for a convergence of the non-perturbative state sums.

The problem with nonperturbative Euclidean quantum gravity is 
two-fold. First, to our knowledge and
aside from our own proposal, no Wick rotation {\it on the space of
all geometries} has ever been constructed in a path-integral context, 
leaving unclear its
relation with the physical theory of Lorentzian gravity. 
Although it is logically possible that a natural Wick rotation
may suggest itself once the Euclidean quantum theory has been solved,
there is so far no convincing evidence (in dimension four) that a 
non-trivial continuum theory exists. This is the second problem of
Euclidean quantum gravity. These conclusions seem to hold for
a wide variety of approaches, including those based on gauge-theoretic 
instead of geometric variables \cite{livrev}. 

Our alternative proposal consists in taking the Lorentzian structure 
seriously from the outset, using a well-defined Wick rotation on
the full, regularized path integral (\ref{pathint}), 
performing all calculations,
simulations and continuum limits in the ``Euclidean sector'', and
finally rotating back the results. As mentioned above, this
program has been carried out successfully and completely explicitly 
in $d\equ 2$ \cite{al}. Another lesson from two-dimensional quantum 
gravity is that the Lorentzian and Euclidean theories will in general be
{\it inequivalent}, even if one can Wick-rotate the final propagator of
the Euclidean theory, in this case, of the Euclidean Liouville theory
(see \cite{india} for an overview of 2d Lorentzian results).
Although we do not wish to exclude the possibility that interesting
theories of Euclidean quantum gravity in higher dimensions exist,
there seems to be a shortage of ideas of how to modify current models
to lead to an interesting continuum behaviour \cite{lattice}.

The apparent inequivalence of our Lorentzian proposal makes it
a promising and genuinely new candidate for a theory of quantum gravity.
More concretely, the geometric constraints which are a natural 
consequence of the causal structure act as a ``regulator'' for
the path integral, suppressing the types of highly degenerate geometries
that seem to be related to the failure to date of the Euclidean 
approaches, be they gauge-theoretic \cite{gaugedegen}, of Regge type 
\cite{reggedegen}, or dynamically triangulated \cite{dtdegen},
to produce any convincing evidence of interesting continuum
physics\footnote{In Euclidean quantum Regge calculus,
an unusual prescription has recently been suggested
to work around the problem of degenerate geometries
at weak gravitational coupling \cite{ha}:
the continuum limit is to be reached by an analytic continuation
from the strong-coupling phase.}.
This issue is discussed further in Sec.\ \ref{kinema} below.

In our construction, we mostly use conventional tools from quantum 
field theory and the theory of critical phenomena 
{\it applied to a diffeomorphism-invariant theory}. 
However, it does not take much to realize that general 
coordinate invariance cannot be accomodated by merely making minor 
changes in the standard QFT formalism. Almost all methods of 
regularizing and renormalizing make use of a fixed metric background 
structure (usually that of flat Minkowski space), albeit often in an
implicit way. By contrast, in a non-perturbative formulation of
quantum gravity the metric degrees of freedom are dynamical and 
initially all on the same footing. A ``ground state of geometry'' should 
only emerge as a {\it solution} to the quantum equations of motion,
much in the same way as a classical physical space-time is the
{\it result} of solving the classical Einstein equations. 

As indicated in (\ref{pathint}),
our path integral will be given as a sum over discretized Lorentzian 
geometries (for given boundary data), 
each weighed with e$^{iS}$, where $S$ is the Regge 
version of the 
classical Lorentzian gravitational action in $d$ space-time dimensions 
($d=3,4$),
\begin{equation}
S [g_{\mu\nu}]=\frac{k}{2} \int_{M} d^dx\; \sqrt{-\det g}\, (R-2\Lambda)
+k \int_{\partial M} d^{d-1}x\; \sqrt{\det h}\, K,
\label{contact}
\end{equation}
with $k^{-1}=8 \pi G_{\rm Newton}$ and the cosmological constant
$\Lambda$. We have included a standard boundary term depending on
the induced metric $h_{ij}$ on the boundary and its extrinsic 
curvature $K_{ij}$. In order to make the state sum well-defined,
we use a non-perturbative Wick rotation on the discretized geometries
to convert the oscillating amplitudes into real Boltzmann factors
e$^{-S^{\rm eucl}}$. It can then be shown that the entire non-perturbative
state sum {\it converges} for suitable choices of the bare coupling
constants, even in the infinite-volume limit. 

It should be emphasized that our construction is intrinsically
{\it quantum}. The formalism of dynamical triangulations is {\it not}
well suited for approximating smooth classical space-times, 
because it lacks the notion of an infinitesimal smooth variation of 
its field variables\footnote{This is different from the situation in 
(quantum) Regge calculus, where continuous changes in 
the edge length variables are allowed.}. 
This would be required if we wanted to
evolve the classical Einstein equations numerically, say. 
It by no means disqualifies the method from use in the quantum
theory. There, one is faced with the problem of having to
approximate ``ergodically'' the space of {\it all} geometries,
whereas in classical applications one is usually interested in
approximating {\it individual}, smooth geometries (often solutions
to the equations of motion). One can compare this situation to
the Feynman path integral for a non-relativistic particle,
which can be obtained as the continuum limit of a sum over
piecewise linear trajectories. 
Although the final state sum is dominated by nowhere 
differentiable (and therefore highly non-classical) paths, 
this is no obstacle to retrieving 
(semi-)classical information about the system from the
expectation values of suitable observables.

Most of the results presented below were already
announced in \cite{ajl}, where also some of the 3d formulas were
given explicitly. Our account here is comprehensive in that we
give all relevant calculations in both three and four dimensions.

We start by a brief description of how to obtain Lorentzian 
space-times by gluing together flat simplicial Minkowskian 
building blocks. We compute their volumes and dihedral angles
as functions of $\alpha$, where $-\alpha$ denotes the squared  
(geodesic) length of one-dimensional time-like edges, and the space-like
edge lengths are fixed to 1. In the next section, we derive
some topological identities among the bulk numbers $N_i$,
$i=1,\ldots, d$ of simplices of dimension $i$. These 
generalize the Euler and Dehn-Sommerville identities to the 
Lorentzian situation where we
distinguish between the space- and time-like character
of the building blocks. These identities are used in the next
section, where the gravitational actions are computed
as functions of $\alpha$, and where we demonstrate how
the corresponding Euclidean actions are obtained by a suitable
analytic continuation of $\alpha$ to negative real values. 

With these ingredients in hand, we define the transfer matrix of
the regularized model in Sec.\ \ref{trans}, together with the 
Hilbert 
space it acts on. Some important properties of the transfer matrix
are proved in Sec.\ \ref{proper}, which ensure the existence of 
a well-defined
Hamiltonian. Paving the way for numerical simulations of the
model, we discuss possible sets of geometric Monte Carlo moves 
in Sec.\ \ref{monte}. 
Using these moves, one can understand certain ``extreme''
regions of the kinematical phase space. We show in Sec.\ \ref{kinema} 
that  the regions of highly
degenerate geometries of Euclidean quantum gravity
cannot be reached by Lorentzian dynamical triangulations.
We end with a summary of current achievements and an 
outlook on the potential applications of discrete 
Lorentzian gravity. The two appendices contain a brief reminder
of some properties of Lorentzian angles and a proof of
link-reflection positivity of 2d Lorentzian gravity.

\section{Discrete Lorentzian space-times}\label{discrete}

Our first task will be to define which discrete Lorentzian
geometries $T$ contribute to the sum in the path integral 
(\ref{pathint}). They are a straightforward generalization of the 
space-time histories used in two dimensions, and have already been 
described \cite{ajl} and used \cite{ajl1} in three dimensions.

Briefly, they can be characterized as ``globally hyperbolic''
$d$-dimensional simplicial manifolds with a sliced structure,
where $(d\mi 1)$-dimensional ``spatial hypersurfaces'' (i.e.
Euclidean $(d\mi 1)$-dimensional equilaterally triangulated manifolds)
of fixed topology are connected by suitable sets of
$d$-dimensional simplices. As a concession to causality, we do not 
allow the spatial slices to change topology. There is a preferred 
notion of a discrete ``time'', namely, the parameter 
labelling successive spatial slices.
Note that this has nothing to do with a gauge choice, since we
were not using coordinates in the first place. This ``proper time''
is simply part of the invariant geometric data common to each of
the Lorentzian geometries. 

We choose particular sets of simplicial building blocks 
in three and four dimensions.
Unlike in the Euclidean case, we cannot make the simplest
possible choice of making all $d$-simplices equilateral. Instead,
we fix all spatial (squared) link lengths to 1,
$l_{\rm space}^{2}\equ 1$, 
and all time-like links to have a squared length $l_{\rm 
time}^{2}\equ -\alpha$, $\alpha >0$. Keeping $\alpha$ variable allows 
for a relative scaling of space- and time-like lengths and is
convenient when discussing the Wick rotation later. 
As usual, the simplices are taken to be pieces of flat
Minkowski space, and a simplicial manifold acquires non-trivial
curvature through the way the individual building blocks are
glued together. For simplicity, we have set the ``lattice spacing''
$a$, which defines a diffeomorphism-invariant cutoff for our model,
equal to 1. (In general, we should have written
$l_{\rm space}^{2}\equ a^{2}$ and $l_{\rm time}^{2}\equ -\alpha a^{2}$.
The correct dependence on $a$ in the formulae below can be
inferred from dimensional considerations.)

As usual in the study of critical phenomena, we expect the final
continuum theory (if it exists) to be largely independent of
the details of the chosen discretization\footnote{In dimension
two, this has already been shown for both the Euclidean \cite{ajm}
and the Lorentzian \cite{lottietal} theories.}. Our choice of building
blocks is simple and allows for a straightforward Wick rotation. 
However, other types of fundamental building blocks may sometimes 
be more convenient. For example, a combination of
pyramids and tetrahedra is used  
in \cite{ajlv} in a three-dimensional context. 
In general, one should be careful not to restrict the local
curvature degrees of freedom too much. With our choice of building 
blocks, the (discretized) local curvatures around given bones
(simplices of dimension $d\mi 2$) can always take both positive and 
negative values.

We will now go on to compute the volumes and dihedral angles of
the $d$-dimen\-sio\-nal Minkowskian simplices, because they are
needed in the gravitational Regge action in dimension $d$. 
All (Lorentzian) volumes in any dimension we will be using
are by definition real and positive. Formulas for Euclidean volumes 
and dihedral angles can be derived from elementary geometric arguments
and may be found in many places in the literature \cite{angles}.
They may be continued to Lorentzian geometries by taking suitable 
care of factors of $i$ and $-1$. We will follow Sorkin's
treatment and conventions for the Lorentzian case \cite{sorkin}.
(Some basic facts about Lorentzian angles are summarized in Appendix 1).
The dihedral angles $\Theta$ are chosen such that $0\leq {\rm Re} 
\Theta \leq \pi$, so giving $\sin\Theta$ and $\cos \Theta$ fixes them 
uniquely. The angles are in general complex, but everything can be 
arranged so that the action comes out real in the end, 
as we shall see.
  
A further ingredient needed
in the calculations are the volumes of lower-dimen\-sio\-nal simplices.
By convention, we have Vol(point)$= 1$. For the one-dimensional
links, 
\begin{equation}
\mbox{ Vol(space-like link)$\ = 1$  \hspace{.4cm} and 
\hspace{.4cm}  Vol(time-like link)$\ =
\sqrt{\alpha }$.}
\end{equation} 
Also for the triangles, we must distinguish between
space- and time-like. The former lie entirely in planes $t\equ const$
whereas the latter extrapolate between two such slices. Their
respective volumes are
\begin{equation}
\mbox{ Vol(space-like triangle)$\ =
\frac{\sqrt{3}}{4}$ \hspace{.4cm} and \hspace{.4cm} 
Vol(time-like triangle)$\ =\frac{1}{4} \sqrt{4 \alpha +1}$.}
\end{equation}

\subsection{Geometry of three-simplices}

The simplices of top-dimension are tetrahedra, of which (up to 
reflection symmetry) there are two types, (3,1) and (2,2). Starting 
with the former (Fig.\ 1a), one computes
\begin{equation}
{\rm Vol}(3,1)=\frac{1}{12}\sqrt{3\alpha +1}.
\end{equation}
\begin{figure}[t]
\centerline{\scalebox{0.6}{\rotatebox{0}
{\includegraphics{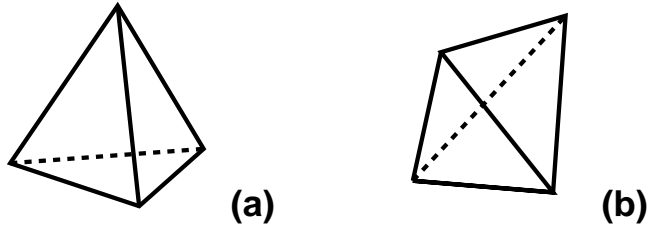}}}}
\caption[3dsimplex]{A (3,1)- and a (2,2)-tetrahedron in three 
dimensions.}
\label{3dsimplex}
\end{figure}
The dihedral angle around any of the space-like bones is given by
\begin{equation}
\cos\Theta_{(3,1)} = -\frac{i}{\sqrt{3} \sqrt{4 \alpha+ 1}},
\hspace{1cm}
\sin\Theta_{(3,1)} = \frac{2 \sqrt{3 \alpha 
+1}}{\sqrt{3} \sqrt{4\alpha+1}},
\label{angle31s}
\end{equation}
and around a time-like bone by
\begin{equation}
\cos\Theta_{(3,1)} = \frac{2\alpha +1}{4 \alpha +1},\hspace{1cm}
\sin\Theta_{(3,1)} = \frac{2 \sqrt{\alpha} \sqrt{3\alpha +1}}{4\alpha +1}.
\label{angle31t}
\end{equation}
For the (2,2)-tetrahedra (Fig.\ 1b) one finds
\begin{equation}
{\rm Vol}(2,2)=\frac{1}{6 \sqrt{2}}\sqrt{2\alpha +1}.
\end{equation}
The dihedral angle around a space-like bone is fixed by
\begin{equation}
\cos\Theta_{(2,2)} = \frac{4 \alpha +3}{4 \alpha 
+1},\hspace{1cm}
\sin\Theta_{(2,2)} = -i\ \frac{2\sqrt{2}\sqrt{2\alpha +1}}{4\alpha+1},
\label{angle22s}
\end{equation}
and that around a time-like bone by
\begin{equation}
\cos\Theta_{(2,2)} = -\frac{1}{4 \alpha+1},\hspace{1cm}
\sin\Theta_{(2,2)} = \frac{2 \sqrt{2 \alpha} \sqrt{2\alpha +1}}{4\alpha +1}.
\label{angle22t}
\end{equation}

\subsection{Geometry of four-simplices}

In $d\equ 4$ there are up to reflection symmetry two types of four-simplices, 
(4,1) and (3,2). In addition to the volumes computed above, we
now also need the volumes of equilateral tetrahedra that lie entirely 
in slices $t\equ const$, given by Vol(space-like 
tetrahedron)$= \frac{1}{6\sqrt{2}}$. For the volume of the 
simplices of type (4,1), Fig.\ 2a, one computes
\begin{equation} 
{\rm Vol}(4,1)= \frac{1}{96} \sqrt{8\alpha +3}.
\end{equation}
\begin{figure}[t]
\centerline{\scalebox{0.6}{\rotatebox{0}
{\includegraphics{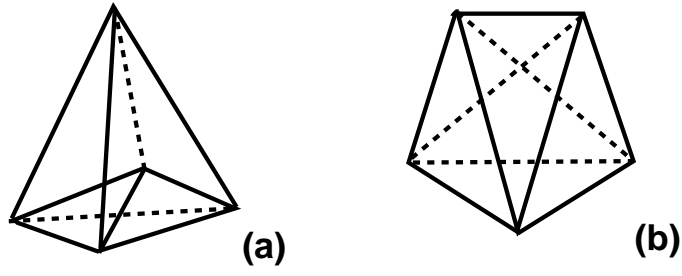}}}}
\caption[4dsimplex]{A (4,1)- and a (3,2)-tetrahedron in four
dimensions.}
\label{4dsimplex}
\end{figure}
A (4,1)-simplex contributes to the curvature with two types of 
dihedral angles (of which there are a total of 10 per simplex), 
depending on whether the parallel transport is taken
around a space- or a time-like triangle. There are six of the former,
each contributing a dihedral angle determined by
\begin{equation}
\cos\Theta_{(4,1)} = -\frac{i}{2\sqrt{2} \sqrt{3\alpha+1}},
\hspace{1cm}
\sin\Theta_{(4,1)} = \sqrt{ \frac{3(8\alpha +3)}{8(3\alpha+1)}},
\label{angle41s}
\end{equation}
and similarly for the angle around time-like triangles,
\begin{equation}
\cos\Theta_{(4,1)} = \frac{2\alpha+1}{2 (3 \alpha+1)},\hspace{1cm}
\sin\Theta_{(4,1)} = \frac{ \sqrt{4\alpha+1} \sqrt{8\alpha +3}}
{2(3\alpha +1)}.
\label{angle41t}
\end{equation}
As in the three-dimensional case (\ref{angle31s}) above, the
trigonometric functions (\ref{angle41s}) are complex, 
with a constant real part 
equal to $\pi/2$, whereas (\ref{angle41t}) for the Euclidean angles 
are real.

The other type of four-simplex is the (3,2)-simplex, see Fig.\ 2b. Its 
volume is given by
\begin{equation}
{\rm Vol}(3,2)=\frac{1}{96} \sqrt{12\alpha +7}.
\end{equation}
As for the dihedral angles, the situation is slightly more involved. Of 
the ten angles contributing per four-simplex there is exactly one 
around a space-like triangle, with
\begin{equation}
\cos\Theta_{(3,2)} = \frac{6 \alpha +5}{2 (3\alpha +1)},\hspace{1cm}
\sin\Theta_{(3,2)} = -i\ \frac{\sqrt{3}\sqrt{12 \alpha +7}}
{2 (3 \alpha+1)},
\label{angle32s}
\end{equation}
which are purely imaginary.
There are two varieties of angles around time-like triangles. If the
angle $\Theta$ is formed by two three-dimensional ``faces'' that are both 
tetrahedra of type (2,2), one has
\begin{equation}
\cos\Theta_{(3,2)} = \frac{4\alpha +3}{4 (2\alpha +1)},\hspace{1cm}
\sin\Theta_{(3,2)} = \frac{\sqrt{(4\alpha+1) (12\alpha +7)}}
{4 (2 \alpha +1)}.
\label{angle32t1}
\end{equation}
By contrast, if the two tetrahedra involved are a pair of a (3,1) and 
a (2,2), the angle between them is defined by
\begin{equation}
\cos\Theta_{(3,2)} = \frac{-1}{2\sqrt{2} \sqrt{2\alpha 
+1}\sqrt{ 3\alpha +1}},\hspace{1cm}
\sin\Theta_{(3,2)} = \frac{\sqrt{(4\alpha+1) (12\alpha +7)}}
{2\sqrt{2} \sqrt{2 \alpha +1}\sqrt{3\alpha +1}}.
\label{angle32t2}
\end{equation}

\section{Topological identities for Lorentzian 
triangulations}\label{topo}

In this section we derive some important linear relations
among the ``bulk'' variables $N_{i}$, $i=0,\ldots,d$ which count the 
numbers of $i$-dimensional simplices in a given 
$d$-dimensional Lorentzian triangulation.
Such identities are familiar from Euclidean dynamically 
triangulated manifolds (see, for example, \cite{aj,mauroetal}). 
The best-known of them is the Euler identity
\begin{equation}
\chi = N_{0}-N_{1}+N_{2}-N_{3}+\ldots,
\end{equation}
for the Euler characteristic $\chi$ of a simplicial manifold with or 
without boundary.
For our purposes, we will need refined versions 
where the simplices are distinguished by their Lorentzian properties.
The origin of these relations lies in the simplicial {\it mani\-fold} 
structure. They can be derived in a systematic way by establishing
relations among simplicial building blocks in local neighbourhoods and 
by summing them over the entire triangulation. Our notation for the
numbers $N_{i}$ is 
\begin{eqnarray}
&&N_{0} = \mbox{ number of vertices}\nonumber\\
&&N_{1}^{\rm TL} = \mbox{ number of time-like links}\nonumber\\
&&N_{1}^{\rm SL} = \mbox{ number of space-like links}\nonumber\\
&&N_{2}^{\rm TL} = \mbox{ number of time-like triangles}\nonumber\\
&&N_{2}^{\rm SL} = \mbox{ number of space-like triangles}\nonumber\\
&&N_{3}^{\rm TL_{1}}\equiv N_{3}^{(3,1)} = 
\mbox{ number of time-like (3,1)- and (1,3)-tetrahedra}\nonumber\\
&&N_{3}^{\rm TL_{2}}\equiv N_{3}^{(2,2)} = 
\mbox{ number of time-like (2,2)-tetrahedra}\nonumber\\
&&N_{3}^{\rm SL} = \mbox{ number of space-like tetrahedra}\nonumber\\
&&N_{4}^{\rm TL_{1}}\equiv N_{4}^{(4,1)} = \mbox{ number of time-like 
(4,1)- and (1,4)-simplices}\nonumber\\
&&N_{4}^{\rm TL_{2}}\equiv N_{4}^{(3,2)} = \mbox{ number of time-like 
(3,2)- and (2,3)-simplices}.
\label{nnumbers}
\end{eqnarray}

\subsection{Identities in 2+1 dimensions}

We will be considering compact spatial slices ${}^{(2)}\Sigma$, 
and either open or periodic boundary conditions in time-direction.
The relevant space-time topologies are therefore $I\times {}^{(2)}\Sigma$ 
(with an initial and a final spatial surface) and 
$S^{1}\times {}^{(2)}\Sigma$. Since the latter results in 
a closed three-manifold, its Euler characteristic vanishes. From this we
derive immediately that  
\begin{equation}
\chi (I\times {}^{(2)}\Sigma)= \chi ({}^{(2)}\Sigma).
\end{equation}
(Recall also that for closed two-manifolds with $g$ handles, we have
$\chi = 2- 2g$, for example, $\chi (S^{2}) =2$ for the two-sphere.) 

Let us for simplicity consider the case of periodic boundary 
conditions. 
A three-dimensional closed triangulation is characterized by the seven 
numbers $N_{0}$, $N_{1}^{\rm SL}$, $N_{1}^{\rm TL}$, $N_{2}^{\rm SL}$, 
$N_{2}^{\rm TL}$, $N_{3}^{(3,1)}$ and $N_{3}^{(2,2)}$. Two relations 
among them are directly inherited from the Euclidean case, namely,
\begin{eqnarray}
&& N_{0} -N_{1}^{\rm SL} -N_{1}^{\rm TL} +N_{2}^{\rm SL}+
N_{2}^{\rm TL} -N_{3}^{(3,1)} -N_{3}^{(2,2)} =0,\label{cons1}\\
&& N_{2}^{\rm SL}+ N_{2}^{\rm TL} = 2 (N_{3}^{(3,1)}+N_{3}^{(2,2)}).
\label{cons2}
\end{eqnarray}
Next, since each space-like triangle is shared by two 
(3,1)-tetrahedra, we have
\begin{equation}
N_{3}^{(3,1)} =\frac{4}{3} N_{1}^{\rm SL}.
\label{31id}
\end{equation}
Lastly, from identities satisfied by the two-dimensional spatial 
slices, one derives
\begin{eqnarray}
&& N_{1}^{\rm SL} =\frac{3}{2} N_{2}^{\rm SL},\\
&& N_{0} =  \chi ({}^{(2)}\Sigma) t+\frac{1}{2} N_{2}^{\rm SL},
\end{eqnarray}
where we have introduced the notation $t$ for the number of 
time-slices in the triangulation. 

We therefore have five linearly independent conditions on the seven 
variables $N_{i}$, leaving us with two ``bulk'' degrees of freedom,
a situation identical to the case of Euclidean dynamical triangulations.
(The variable $t$ does not have the same status as the 
$N_{i}$, since it scales (canonically) only like a length, and not like a
volume.) 

\subsection{Identities in 3+1 dimensions}

Here we are interested in four-manifolds which are of the form of a 
product of a compact three-manifold ${}^{(3)}\Sigma$ with either an
open interval or a circle, that is 
$I\times {}^{(3)}\Sigma$ or $S^{1}\times {}^{(3)}\Sigma$. (Note that 
because of $\chi ({}^{(3)}\Sigma)=0$, we have
$\chi (I\times {}^{(3)}\Sigma) =\chi (S^{1}\times {}^{(3)}\Sigma)$.
An example is $\chi (S^{1}\times T^{3})\equiv \chi (T^{4})=0$.)
In four dimensions, we need the entire set (\ref{nnumbers}) of ten 
bulk variables $N_{i}$. Let us again discuss the linear constraints 
among them for the case of periodic boundary conditions in time.

There are three constraints which are inherited from the 
Dehn-Sommerville conditions for general four-dimensional triangulations
\cite{aj,mauroetal},
\begin{eqnarray}
&&N_{0} -N_{1}^{\rm SL} -N_{1}^{\rm TL} +N_{2}^{\rm SL} +N_{2}^{\rm TL} 
- N_{3}^{\rm SL}- N_{3}^{\rm TL_{1}} -
N_{3}^{\rm TL_{2}}+ N_{4}^{\rm TL_{1}}+ N_{4}^{\rm TL_{2}}=\chi,
\nonumber\\
&&2 (N_{1}^{\rm SL}\pl N_{1}^{\rm TL}) -3 (N_{2}^{\rm SL}\pl N_{2}^{\rm 
TL}) + 4(N_{3}^{\rm SL}\pl N_{3}^{\rm TL_{1}}\pl N_{3}^{\rm TL_{2}}) -
5 (N_{4}^{\rm TL_{1}}\pl N_{4}^{\rm TL_{2}}) =0,\nonumber\\
&& 5 (N_{4}^{\rm TL_{1}}+ N_{4}^{\rm TL_{2}}) = 2
(N_{3}^{\rm SL}+ N_{3}^{\rm TL_{1}} +N_{3}^{\rm TL_{2}}).
\end{eqnarray}
The remaining constraints are special to the sliced, Lorentzian
space-times we are using. There are two which arise from conditions on 
the space-like geometries alone (cf. (\ref{cons1}), (\ref{cons2})),
\begin{eqnarray}
&&N_{0}- N_{1}^{\rm SL} +N_{2}^{\rm SL} -N_{3}^{\rm SL} =0,\nonumber\\
&&N_{2}^{\rm SL}=2 N_{3}^{\rm SL}.
\end{eqnarray}
Furthermore, since each space-like tetrahedron is shared by a pair of a 
(4,1)- and a (1,4)-simplex,
\begin{equation}
2 N_{3}^{\rm SL} = N_{4}^{(4,1)},
\end{equation}
and since each time-like tetrahedron of type (2,2) is shared by a
pair of (3,2)-simplices, we have
\begin{equation}
2 N_{3}^{\rm TL_{2}} = 3 N_{4}^{(3,2)}.
\end{equation}

In total, these are seven constraints for ten variables. This is to 
be contrasted with the case of Euclidean triangulations, where there 
are only two bulk variables. With the help of the results of 
Sec.\ \ref{kinema} 
below, one can convince oneself that there are
no further linear constraints in the Lorentzian case. 
Taking $N_{2}^{\rm TL}$, $N_{4}^{(4,1)}$ and 
$N_{4}^{(3,2)}$ as the three remaining variables, it is 
straightforward to verify that a relation of the form
\begin{equation}
\alpha N_{2}^{\rm TL} +\beta N_{4}^{(4,1)} +\gamma N_{4}^{(3,2)} 
+\delta = 0
\end{equation}
is not compatible with the Monte Carlo moves, 
(\ref{delf41})-(\ref{delf44}), unless 
$\alpha\equ\beta\equ\gamma\equ\delta\equ 0$.

\section{Actions and the Wick rotation}\label{actions}

We are now ready to construct the gravitational 
actions of Lorentzian dynami\-cal triangulations explicitly.
We will investigate their behaviour in the complex
$\alpha$-plane and show that -- subject to a
dimension-dependent lower bound on $\alpha$ -- the map
\begin{equation}
\alpha \mapsto -\alpha,\;\;\;\;\;\;\;\;\;\;\; 
\alpha >\frac{1}{2}\;\; (d=3),
\;\;\;\;
\alpha > \frac{7}{12}\;\; (d=4),
\label{wick}
\end{equation}
where $\alpha$ is continued through the lower half of the complex 
plane, defines a non-perturbative Wick rotation from Lorentzian to
Euclidean discrete geometries. Under this map, the weights in
the partition function transform according to
\begin{equation}
{\rm e}^{iS}\mapsto {\rm e}^{-S^{\rm eucl}}.
\end{equation}
For the special case
$\alpha =-1$, the usual expressions for the actions used in
equilateral Euclidean dynamical triangulations are reproduced.

\subsection{Gravitational action in three dimensions}

The Regge analogue of the continuum action (\ref{contact}) in $d\equ 3$ is
given by (c.f. \cite{sorkin})
\begin{eqnarray}
S^{(3)}&=&k\sum_{\stackrel{\rm space-like}{l}}{\rm Vol}(l)\ \frac{1}{i}
(2\pi -\sum_{\stackrel{\rm tetrahedra}{{\rm at}\ l}}\Theta)+
k\sum_{\stackrel{\rm time-like}{l}}{\rm Vol}(l)\  
(2\pi -\sum_{\stackrel{\rm tetrahedra}{{\rm at}\ l}}\Theta)\nonumber\\
&&-\lambda \sum_{\stackrel{\rm (3,1)\& (1,3)-}{\rm tetrahedra}}{\rm Vol}(3,1) -
\lambda \sum_{\stackrel{\rm (2,2)-}{\rm tetrahedra}}{\rm Vol}(2,2),
\label{act3dis}
\end{eqnarray}
where we have now set $\lambda =k\Lambda$.
Performing the sums, and taking into account how many tetrahedra meet
at the individual links, one can re-express the action as a function 
of the bulk variables $N_{1}$ and $N_{3}$, namely,
\begin{eqnarray}
S^{(3)}\!\! &=&\!\! k\left( \frac{2\pi}{i} N_{1}^{\rm SL}-\frac{2}{i} 
N_{3}^{(2,2)} \arcsin\frac{-i\ 2\sqrt{2}\sqrt{2\alpha +1}}{4\alpha 
+1}- \frac{3}{i}  
N_{3}^{(3,1)} \arccos\frac{-i}{\sqrt{3}\sqrt{4\alpha +1}}
\right)\nonumber\\
&&+k\sqrt{\alpha} \left( 2\pi N_{1}^{\rm TL}-4 N_{3}^{(2,2)} 
\arccos \frac{-1}{4\alpha +1} -3 N_{3}^{(3,1)} \arccos 
\frac{2\alpha+1}{4\alpha+1} \right)\nonumber\\
&&-\lambda \left( N_{3}^{(2,2)}\ \frac{1}{12} \sqrt{ 4\alpha 
+2}+N_{3}^{(3,1)}\ \frac{1}{12} \sqrt{3\alpha +1} \right).
\label{act3disN}
\end{eqnarray}
Our choice for the inverse trigonometric functions with imaginary
argument from (\ref{angle31s}), (\ref{angle22s}), avoids 
branch-cut ambiguities for real, positive $\alpha$.
Despite its appearance, the action (\ref{act3disN}) 
is {\it real} in the relevant
range $\alpha >0$, as can be seen by applying elementary 
trigonometric identities and the relation (\ref{31id}). 
The final result for the Lorentzian action can be written as a
function of three bulk variables (c.f. Sec.\ \ref{topo}), for example,
$N_{1}^{\rm TL}$, $N_{3}^{(3,1)}$ and $N_{3}^{(2,2)}$, as
\begin{eqnarray}
S^{(3)}&=& 2\pi k\sqrt{\alpha} N_{1}^{\rm TL} \nonumber\\
&+&N_{3}^{(3,1)} 
\Bigl( -3 k\ {\rm arcsinh}\ \frac{1}{\sqrt{3}\sqrt{4\alpha +1}}
-3 k \sqrt{\alpha} \arccos \frac{2\alpha+1}{4\alpha +1} -
\frac{\lambda}{12} \sqrt{3\alpha +1}\Bigr)\nonumber \\
&+&N_{3}^{(2,2)}
\Bigl( 2 k\ {\rm arcsinh}\ \frac{2\sqrt{2}\sqrt{2\alpha +
1}}{4\alpha +1}
-4 k \sqrt{\alpha} \arccos \frac{-1}{4\alpha+1} -
\frac{\lambda}{12} \sqrt{ 4\alpha +2} \Bigr).\nonumber \\
\label{3dloract}
\end{eqnarray}

Since we are interested in continuing this action to the Euclidean
sector where $\alpha$ is real and negative, we
need to understand the properties of (\ref{3dloract}) in the complex
$\alpha$-plane. For general $\alpha$ (and assuming $k$ and $\lambda$
real and arbitrary), $S^{(3)}(\alpha )$ takes complex values.
There are two special cases: (i) for $\alpha \in\R$, $\alpha \geq 0$,
the entire action is real -- 
this is simply the Lorentzian case considered above;
(ii) for $\alpha\in\R$, $\alpha\leq -\frac{1}{2}$, the expression 
(\ref{3dloract}) is purely imaginary and we will see that it
coincides with $i S^{\rm eucl}$. The reason for the non-trivial upper 
bound of $-\frac{1}{2}$ are the triangle inequalities for Euclidean
signature. It is easy to verify that as the positive Euclidean squared 
length $-\alpha$ becomes shorter, the (2,2)-tetrahedral building blocks 
degenerate first, at $-\alpha =\frac{1}{2}$, followed by the 
(3,1)-tetrahedra at $-\alpha =\frac{1}{3}$. 
Therefore, if for fixed $\alpha$ we adopt $\alpha\mapsto -\alpha$ as
prescription for a Wick rotation, we should confine ourselves to
values $|\alpha | >\frac{1}{2}$. 

Because of the appearance of
square roots in the action (\ref{3dloract}), there is a two-fold 
ambiguity in what we mean by $S^{(3)}(-\alpha )$. It turns out that
the correct Euclidean expression is obtained by continuing $\alpha$
through the lower-half complex plane such that the branch cuts along
the negative real axis are approached from below. In other words,
if the argument of an expression of the form $\sqrt{\alpha +c}$
becomes negative as a result of the Wick rotation, we interpret it
as $-i\sqrt{-(\alpha +c)}$. 

Let us now consider the special case $\alpha =1$, together with its
Wick rotation. The Lorentzian action becomes 
\begin{eqnarray}
S^{(3)}(\alpha =1)&= & 2 \pi k N_1^{\rm TL} +
N_3^{(3,1)} (-3.548 k -0.167 \lambda) +
N_3^{(2,2)} (-5.355 k -0.204 \lambda)\nonumber\\
&=& 2 \pi k \chi T +
N_3^{(3,1)} (-0.407 k -0.167 \lambda) +
N_3^{(2,2)} (0.929 k -0.204 \lambda),\nonumber\\
\end{eqnarray}
where in the last step we have used the Lorentzian manifold 
identities derived in Sec.\ \ref{topo}.
Its Wick-rotated Euclidean counterpart is given by
\begin{eqnarray}
S^{(3)}(\alpha =\mi 1)\!\! &=&\!\! -2 \pi i k N_1^{\rm TL} +
N_3^{(3,1)} (2.673 i k +0.118 i\lambda) +
N_3^{(2,2)} (7.386 i k +0.118 i\lambda)\nonumber\\
&=& \!\! -2 \pi i k (N_1^{\rm TL}+N_1^{\rm SL}) +
(N_3^{(3,1)} +N_3^{(2,2)}) (7.386 i k +0.118 i\lambda)\nonumber\\
&\equiv &\!\! -2 \pi i k N_1 +N_3 (6 i k \arccos \frac{1}{3} +
\frac{i}{6\sqrt{2}}\lambda )\equiv iS_{\rm EDT}^{(3)},
\label{3dactfinal}
\end{eqnarray}
where in the last step we have identified the well-known expression
for the action of Euclidean dynamically triangulated gravity in
three dimensions.

\subsection{Gravitational action in four dimensions}

The form of the discrete action in four dimensions is completely 
analogous to (\ref{act3dis}), that is,
\begin{eqnarray}
S^{(4)}&=&k\sum_{\stackrel{\rm space-like}{ \triangle}}{\rm Vol}(\triangle) 
\ \frac{1}{i}
(2\pi -\sum_{\stackrel{\rm 4-simplices}{{\rm at}\ \triangle}}\Theta)+
k\sum_{\stackrel{\rm time-like}{\triangle}}{\rm Vol}(\triangle)\  
(2\pi -\sum_{\stackrel{\rm 4-simplices}{{\rm at}\ \triangle}}\Theta)\nonumber\\
&&-\lambda  \sum_{\stackrel{\rm (4,1)\& (1,4)-}{\rm tetrahedra}}
{\rm Vol}(4,1) -
\lambda \sum_{\stackrel{\rm (3,2) \& (2,3)-}{\rm tetrahedra}}{\rm Vol}(3,2).
\label{act4dis}
\end{eqnarray}
Expressed in terms of the bulk variables $N_{2}$ and $N_{4}$, the 
action reads
\begin{eqnarray}
&&S^{(4)}=k \biggl( \frac{2\pi}{i} \frac{\sqrt{3}}{4} N_{2}^{\rm SL} -
\frac{\sqrt{3}}{4i} N_{4}^{(3,2)} \arcsin\frac{-i\ \sqrt{3}\sqrt{12 \alpha +7}}
{2 (3 \alpha+1)} \nonumber\\
&& - \frac{\sqrt{3}}{i} N_{4}^{(4,1)} 
\arccos\frac{-i}{2\sqrt{2}\sqrt{3\alpha +1}} \biggr)
+ \frac{k}{4} \sqrt{4\alpha +1}\ 
\Biggl( 2\pi N_{2}^{\rm TL} -\nonumber\\
&&-N_{4}^{(3,2)} \Bigl( 
6\arccos\frac{-1}{2\sqrt{2}\sqrt{2\alpha +1}\sqrt{3\alpha +1}} +
3\arccos\frac{4\alpha +3}{4 (2\alpha +1)} \Bigr) \nonumber\\
&&-6 N_{4}^{(4,1)} \arccos\frac{2\alpha +1}{2 (3\alpha +1)}\Biggr)
-\lambda \Bigl( N_{4}^{(4,1)}\ \frac{\sqrt{8\alpha +3}}{96} +
N_{4}^{(3,2)}\ \frac{\sqrt{12\alpha +7}}{96} \Bigr) .
\label{act4disN}
\end{eqnarray}
We have again taken care in choosing the inverse functions of the
Lorentzian angles in (\ref{angle41s}) and (\ref{angle32s}) that
make the expression (\ref{act4disN}) unambiguous. Using the manifold 
identities for four-dimensional simplicial Lorentzian triangulations 
derived in Sec.\ \ref{topo}, 
the action can be rewritten as a function of the three bulk 
variables $N_{2}^{\rm TL}$, $N_{4}^{(3,2)}$ and $N_{4}^{(4,1)}$, in a 
way that makes its real nature explicit,
\begin{eqnarray}
S^{(4)}\!\! &=&\!\! \frac{k \pi}{2} \sqrt{4\alpha +1}\ N_{2}^{\rm TL} 
+ N_{4}^{(4,1)}\cdot \nonumber\\
&&
\!\!\Biggl( -\sqrt{3} k\ {\rm arcsinh} \frac{1}{2\sqrt{2}\sqrt{3\alpha +1}}
-\frac{3 k}{2} \sqrt{4\alpha +1} \arccos\frac{2\alpha +1}{2 (3\alpha +1)}
-\lambda \frac{\sqrt{8\alpha +3}}{96}
\Biggr)\nonumber \\
&+&\!\! N_{4}^{(3,2)}
\Biggl( \frac{\sqrt{3} k}{4}\ {\rm arcsinh}\frac{\sqrt{3}\sqrt{12 \alpha +7}}
{2 (3 \alpha+1)} 
- \frac{3 k}{4} \sqrt{4\alpha +1}\nonumber\\
&&\!\!\biggl( 2\arccos\frac{-1}{2\sqrt{2}\sqrt{2\alpha +1}\sqrt{3\alpha +1}} +
\arccos\frac{4\alpha +3}{4 (2\alpha +1)}\biggr) 
-\lambda \frac{\sqrt{12\alpha +7}}{96}\Biggr).
\label{4dloract}
\end{eqnarray}
It is straightforward to verify that this action is real for real
$\alpha \geq -\frac{1}{4}$, and purely imaginary for $\alpha\in\R$, 
$\alpha \leq -\frac{7}{12}$. Note that this implies that we could
in the Lorentzian case choose to work with building blocks possessing
light-like (null) edges ($\alpha =0$) instead of time-like edges, or even
work entirely with building blocks whose edges are all space-like.
However, if we want to stick to our simple prescription 
(\ref{wick}) for the Wick rotation, we should choose only values
$|\alpha | >\frac{7}{12}$. Only in this case will we avoid 
degeneracies in the Euclidean regime in the form of violations of
triangle inequalities. (Coming from large positive length-squared
$-\alpha$, it is the (3,2)-simplices that degenerate first, at
$-\alpha =\frac{7}{12}$, followed by the (4,1)-simplices at
$-\alpha =\frac{3}{8}$.) 

As in three dimensions, in order to obtain the Euclidean version of 
the action, we continue the square roots in (\ref{4dloract}) in the 
lower half of the complex $\alpha$-plane. Let us again consider the
two special cases $\alpha =\pm 1$. Evaluating (\ref{4dloract})
numerically for the Lorentzian case, we obtain
\begin{eqnarray}
S^{(4)}(\alpha \equ 1) &\!\equ\! & 3.512 k N_{2}^{\rm TL} + N_{4}^{(4,1)} 
(-4.284 k -0.035 \lambda) 
+N_{4}^{(3,2)} (-6.837 k -0.045\lambda )\nonumber\\
&\equiv & 7.025 k N_{0} + N_{4}^{(4,1)} 
(-0.772 k -0.035 \lambda) 
+N_{4}^{(3,2)} (0.188 k -0.045\lambda ),\nonumber\\
\end{eqnarray}
where we have used some topological identities from Sec.\ \ref{topo}.
Its Wick-rotated version is given by
\begin{eqnarray}
S^{(4)}(\alpha \equ -1)&\equ & 
-2.721 i k N_2^{\rm TL} +
N_4^{(4,1)} (2.987 i k +0.023 i\lambda) +\nonumber\\
&& \hspace{.5cm} N_4^{(3,2)} (5.708 i k +0.023 i\lambda)\nonumber\\
&=& 
-2.721 i k (N_2^{\rm TL} +N_{2}^{\rm SL}) +
(N_4^{(4,1)} +N_4^{(3,2)}) (5.708 i k +0.023 i\lambda)\nonumber\\
&\equiv & -\frac{\sqrt{3}}{2}\pi i k N_{2} +N_{4} 
( \frac{5 \sqrt{3}}{2} i k \arccos \frac{1}{4} +
\frac{i\sqrt{5}}{96} \lambda) \equiv iS_{\rm EDT}^{(4)},
\label{4dactfinal}
\end{eqnarray}
where the expression in the last line is recognized as the
standard form of the action for Euclidean dynamical triangulations in
four dimensions.

\section{The transfer matrix}\label{trans}

We have now all the necessary prerequisites to study the full
partition function or amplitude for pure gravity in the
dynamically triangulated model,
\begin{equation}
Z=\sum_{\mbox{triangulations T}} \frac{1}{C(T)}\ {\rm e}^{iS(T)},
\label{part}
\end{equation}
where the sum is taken over inequivalent Lorentzian triangulations, 
corresponding to inequivalent discretized space-time geometries. 
We have chosen the measure factor to have the standard form
in dynamical triangulations, $m(T)=1/C(T)$. 
The combinatorial weight of each
geometry is therefore not 1, but the inverse of the order $C(T)$ 
of the symmetry (or automorphism) group of the triangulation $T$. One
may think of this factor as the remnant of the division by the volume 
of the diffeomorphism group Diff($M$) that would occur in a formal 
gauge-fixed
continuum expression for $Z$. Its effect is to suppress geometries
possessing special symmetries. This is analogous to what
happens in the continuum where the diffeomorphism orbits through
metrics with special isometries are smaller than the ``typical'' orbits,
and are therefore of measure zero in the quotient space
Metrics/Diff$(M)$. 

Again, we expect the final result to be independent of the details
of the dynamical and combinatorial weights in (\ref{part}). The 
sparseness of critical points and universality usually ensure that
the same continuum limit (if existent) is obtained for a wide variety 
of initial regularized models.

The most natural boundary conditions for $Z$ in our discretized
Lorentzian model are given by specifying spatial discretized
geometries $g$ at given initial and finite proper times $t$. 
The slices at constant $t$ are by construction space-like and of fixed
topology,
and are given by (unlabelled) {\it Euclidean} triangulations
in terms of $(d-1)$-dimensional building blocks (equilateral
Euclidean triangles in $d=3$ and equilateral Euclidean tetrahedra
in $d=4$). 

Given an initial and a final geometry $g_{1}$ and $g_{2}$, 
we may think of the corresponding amplitude $Z$ as the matrix element 
of the quantum propagator of the system, evaluated between two
states $|g_{1}\rangle$ and $|g_{2}\rangle$. Since we regard
distinct spatial triangulations $g_{i}$ as physically
inequivalent, a natural scalar product is given by
\begin{equation}
\la g_1 | g_2 \ra = \frac{1}{C(g_1)} \delta_{g_1,g_2}, ~~~~
\sum_g C(g) \; |g\ra \la g | = \hat{1}.
\label{norm}
\end{equation}
In line with our previous reasoning, we have included a symmetry
factor $C(g)$ for the spatial triangulations. 

In the regularized context, it is natural to have a cutoff on the 
allowed size of the spatial slices so that their volume is
$\leq N$. The spatial discrete volume Vol($g$) is simply the number of
$(d-1)$-simplices in a slice of constant integer $t$. 
We define the finite-dimensional Hilbert space 
$H^{(N)}$ as the space spanned by
the vectors $\{ |g\rangle,\ N_{\rm min}\leq {\rm Vol}(g) \leq N \}$, 
endowed with the scalar product (\ref{norm}). The lower bound
$N_{\rm min}$ is the minimal size of a spatial triangulation of
the given topology satisfying the simplicial manifold conditions.
It is well-known that 
the number of states in $H^{(N)}$ is exponentially bounded as a 
function of $N$ \cite{av}.

For given volume-cutoff $N$, we can now associate 
with each time-step  $\Delta t \equ 1$ a transfer matrix
${\hat T}_N$ describing the evolution of the system from $t$ to $t\pl 1$,
with matrix elements  
\beq{transfer}
\langle g_2|\hat T_N(\alpha ) |g_1\rangle\equiv G_\alpha (g_1,g_2;1)
=\sum_{T:\ g_1\rightarrow g_2} \frac{1}{C(T)}\  
\e^{i \Delta S_\alpha (T)}.
\eeq
The sum is taken over all distinct interpolating $d$-dimensional
triangulations $g_{1}\rightarrow g_{2}$ 
of the ``sandwich" with boundary geometries $g_1$
and $g_2$, contributing $\Delta S$ 
to the action, according to (\ref{3dloract}), 
(\ref{4dloract}). 
The propagator $G_N(g_1,g_2;t)$ for arbitrary time intervals
$t$ is obtained by iterating the transfer matrix $t$ times,
\beq{prop}
G_N(g_1,g_2;t)=\langle g_2|\hat T_N^t|g_1\rangle,
\eeq
and satisfies the semigroup property
\beq{iter}
G_N(g_1,g_2;t_1+t_2)=\sum_g C(g)\  G_N(g_{1},g;t_1)\ G_N(g,g_{2};t_2),
\eeq
where the sum is over all spatial geometries (of bounded volume) at 
some intermediate time.
Because of the appearance of {\it different} symmetry factors in 
(\ref{transfer}) and (\ref{iter}) it is at first not obvious why
the composition property (\ref{iter}) should hold. 
In order to understand that it does, one has to realize that
by continuity and by virtue of the manifold property there are no 
non-trivial automorphisms of a sandwich $T =g_{1}\rightarrow g_{2}$ that 
leave its boundary invariant. It immediately follows that the (finite)
automorphism group of $T$ must be a subgroup
of the automorphism groups of both of the boundaries, and that
therefore $C(T)$ must be a divisor of both $C(g_{1})$ and $C(g_{2})$. 
It is then straightforward to verify that the factor $C(g)$
appearing in the composition law (\ref{iter}) ensures that 
the resulting geometries appear exactly with the symmetry
factor they should have according to (\ref{transfer}).

Our construction of the propagator obeying (\ref{iter}) 
parallels the two-dimensional treatment of \cite{al}. 
Also in higher dimensions one may use the trick of introducing 
``markings'' on the 
boundary geometries to absorb some of the symmetry factors $C$.
However, note that unlike in $d=2$ the number of possible locations 
for the marking does not in general 
coincide with the order $C(g)$ of the boundary's symmetry group.

\section{Properties of the transfer matrix}\label{proper}

In this section we will establish some desirable properties of
the transfer matrix $\hat T_{N}(\alpha)$, defined in (\ref{transfer}),
which will enable us to define a self-adjoint Hamiltonian operator.
Since the name ``transfer matrix'' in statistical mechanical models
is usually reserved for the Euclidean expression
$\hat T={\rm e}^{-a\hat h}$, where $a$ denotes the lattice spacing in 
time-direction, we will from now on work with
the Wick-rotated matrix $\hat T_{N}(-1)$. 
A set of sufficient conditions on $\hat T_{N}$ 
guaranteeing the existence of a 
well-defined quantum Hamiltonian $\hat h$ are that the transfer
matrix should satisfy
\begin{itemize}
\item[(a)] {\it symmetry}, that is, $\hat T_{N}^{\dagger}=\hat T_{N}$.
This is the same as self-adjointness since the Hilbert space $H^{(N)}$
which $\hat T_{N}$ acts on is finite-dimensional. It is necessary if the 
Hamiltonian is to have real eigenvalues. 
\item[(b)] {\it Strict positivity} is required in addition to (a), 
that is, all eigenvalues must be greater than zero; otherwise,
$\hat h_{N}=-a^{-1} \log \hat T_{N}$ does not exist.
\item[(c)] {\it Boundedness}, that is, in addition to (a) and (b), 
$\hat T_{N}$ should be bounded above 
to ensure that the eigenvalues of the Hamiltonian are bounded below, 
thus defining a stable physical system. 
\end{itemize} 
Establishing (a)-(c) suffices to show that our discretized systems are 
well-defined as regularized statistical models. This of course does
not imply that they will possess interesting continuum limits and that
these properties will necessarily persist in the limit (as would be desirable).
On the other hand, it is difficult to imagine how the continuum limit 
could have such properties unless also the limiting sequence of
regularized models did.

All of the above properties are indeed satisfied for the Lorentzian 
model in $d\equ 2$ \cite{al}, where moreover the quantum Hamiltonian and 
its complete spectrum in the continuum limit are known explicitly 
\cite{nakayama,lottietal}. Note that self-adjointness of the continuum 
Hamiltonian $\hat H$ implies a unitary time evolution operator
${\rm e}^{-i\hat H T}$
if the continuum proper time $T$ is analytically continued.
In $d>2$ we are able to prove a slightly weaker statement than
the above, namely, we can verify (a)-(c) for the
two-step transfer matrix $\hat T_N^2$.\footnote{This corrects 
a claim in \cite{ajl} where we stated that (b) holds for
the one-step transfer matrix.} This is still
sufficient to guarantee the existence of a well-defined Hamiltonian.

One verifies the symmetry of the transfer matrix by inspection of 
the explicit form of the matrix elements (\ref{transfer}). The
``sandwich actions'' $\Delta S_{\alpha}$ as functions of the boundary 
geometries $g_{1}$, $g_{2}$ in three and four dimensions can be read 
off from (\ref{3dloract}) and (\ref{4dloract}). 
To make the symmetry explicit, one may simply rewrite these 
actions as separate functions of the simplicial building blocks and
their mirror images under time-reflection (in case they are
different). Likewise, the symmetry factor $C(T)$ and the counting
of interpolating geometries in the sum over $T$ are invariant
under exchange of the in- and out-states, $|g_{1}\rangle$ and
$|g_{2}\rangle$.

Next, we will discuss the 
reflection (or Osterwalder-Schrader) positivity 
\cite{oschrader,glimmja} of our model, 
with respect to reflection at planes of constant integer and
half-integer time (see also \cite{momu} and references therein). 
These notions 
can be defined in a straightforward way in the Lorentzian model
because it possesses a distinguished notion of (discrete proper) time.
Reflection positivity implies the {\it positivity} of the transfer 
matrix, $\hat T_{N}\geq 0$. 

``Site reflection'' denotes the reflection $\theta_{s}$ 
with respect to a spatial 
hypersurface of constant integer-$t$ (containing ``sites'', i.e. 
vertices), for example, $t\equ 0$, where it takes the form
\begin{equation}
\theta_{s}:\ t\rightarrow -t.
\label{sreflect}
\end{equation}
Let us accordingly split any triangulation $T$ along this hypersurface, 
so that $T^{-}$ is the triangulation with $t\leq 0$ and $T^{+}$ the one 
with $t\geq 0$, and $T^{-} \cap T^{+} = g(t\equ 0)\equiv g_{0}$, where 
$g_{0}$ denotes a spatial triangulation at $t\equ 0$. 
Consider now functions $F$ that depend only on $T^{+}$ (that is,
on all the connectivity data specifying $T^{+}$ uniquely, in some 
parametrization of our choice). Site-reflection positivity means the
positivity of the Euclidean expectation value
\begin{equation}
\langle (\theta_{s}F) F\rangle \geq 0,
\label{fspos}
\end{equation}
for all such functions $F$. The action of $\theta_{s}$ on a function
$F(T^+)$ is defined by anti-linearity, 
$(\theta_{s}F)(T^-):= \bar F(\theta_{s}(T^-))$. 
By virtue of the composition property (\ref{iter}), we can write
\begin{eqnarray}
\langle (\theta_{s}F) F\rangle &=& Z^{-1} \sum_{T} \frac{1}{C(T)} 
(\theta_{s}F) F\ {\rm e}^{-S(T)}\nonumber\\
& = &
Z^{-1} \sum_{g_{0}} C(g_{0}) \sum_{ 
{ T^{-}\atop T^{-}(t=0)=g_{0} } }
\frac{(\theta_{s}F)(T^{-})}{C(T^{-})}\ {\rm e}^{-S(T^{-})}
\sum_{{ T^{+}\atop T^{+}(t=0)=g_{0} } }
\frac{F(T^{+})}{C(T^{+})}\ {\rm e}^{-S(T^{+})}\nonumber\\
&=& Z^{-1} \sum_{g_{0}} C(g_{0}) \bar{\cal F}(g_{0}) {\cal F}(g_{0})
\geq 0.
\label{scal}
\end{eqnarray}
The equality in going to the last line 
holds because both the action and the symmetry factor $C$ 
depend on ``outgoing'' and ``incoming''
data in the same way (for example, on (m,n)-simplices in the same way 
as on (n,m)-simplices). Note that the standard procedure of
extracting a scalar product and a Hilbert space from (\ref{scal}) 
is consistent with our earlier definition (\ref{norm}). One
associates (in a many-to-one fashion) functions $F(T^{+})$ with
elements $\Psi_{F}$ of a Hilbert space at fixed time $t$ with
scalar product $\langle\cdot ,\cdot\rangle$, where 
$\langle \Psi_{F},\Psi_{G}\rangle =\langle (\theta_{s}F)G\rangle$
\cite{oschrader}. A set of representative functions which 
reproduces the states and orthogonality relations (\ref{norm}) is
given by
\begin{equation}
F_{g}(T^{+})=\left\{ {1/\sqrt{C(g)},\;\;\; T^{+}(t=0)=g\atop
0\;\;\;\;\;\;\;\;\;\;\; {\rm otherwise,} } \right.
\end{equation}
as can be verified by explicitly computing the expectation
values $\langle (\theta_{s}F_{g'})F_{g''}\rangle$. 

We have therefore proved site-reflection positivity of our model.
This is already enough to construct a Hamiltonian from 
the square of the transfer matrix $\hat T_{N}$ (see eq.\ (\ref{ham2})
below), since it implies the positivity of $\hat T_{N}^{2}$ \cite{momu}. 

Proving in addition link-reflection positivity (which would imply
positivity of the ``elementary'' transfer matrix, and not only of
$\hat T_{N}^{2}$) turns out to be more involved.
A ``link reflection'' is the reflection 
$\theta_{l}$ at a hypersurface of
constant half-integer-$t$, for example, $t\equ 1/2$,
\begin{equation}
\theta_{l}:\ t\rightarrow 1-t.
\label{lreflect}
\end{equation} 
To show link-reflection positivity in our model we would need
to demonstrate that
\begin{equation}
\langle (\theta_{l}F) F\rangle \geq 0,
\label{flpos}
\end{equation}
where $F$ is now any function that depends only on the part
$T^+$ of the triangulation $T$ at times later or equal to 1.
We can again write down the expectation value,
\begin{eqnarray}
\langle (\theta_{l}F) F\rangle &=& Z^{-1} \sum_{g_0} \sum_{g_1}
C(g_0) C(g_1) G_N(g_0,g_1;1)\times\nonumber\\
&& \hspace{.3cm} \sum_{ { T^{-}\atop T^{-}(t=0)=g_{0} } }
\frac{(\theta_{s}F)(T^{-})}{C(T^{-})}\ {\rm e}^{-S(T^{-})}
\sum_{{ T^{+}\atop T^{+}(t=1)=g_{1} } }
\frac{F(T^{+})}{C(T^{+})}\ {\rm e}^{-S(T^{+})}.
\label{linkref}
\end{eqnarray}
In order to show that this is positive, one should try to
rewrite the right-hand side as a sum of positive terms. 
The proof of this is
straightforward in $d=2$ (see App.\ 2 for details), 
but it is considerably more
difficult to understand what happens in higher dimensions.
The reason for this is the non-trivial way in which the various
types of simplicial building blocks fit together in between
slices of constant integer-$t$. In a way, this is a desirable
situation since it means that there is a more complicated
``interaction'' among the simplices. It is perfectly possible
that $\hat T_{N}$ itself is {\it not} positive for 
$d=3,4$. This may depend both on the values of the bare
couplings and on the detailed choices we have made as part of 
our discretization.
(By contrast, it is clear from our proof of site-reflection positivity
that this property is largely independent of the choice of building 
blocks.)

Nevertheless, as already mentioned above, site-reflection positivity 
is perfectly sufficient for the construction of a well-defined
Hamiltonian. So far, we have only shown that the eigenvalues of the
squared transfer matrix are positive. In order for the
Hamiltonian
\begin{equation}
\hat h'_N := -\frac{1}{2a} \log \hat T_N^2
\label{ham2}
\end{equation}
to exist, we must achieve strict positivity. 
We do not expect that the Hilbert space $H^{(N)}$ contains any
zero-eigenvectors of $\hat T_N$
since this would entail a ``hidden''
symmetry of the discretized theory. It is straightforward to
see that none of the basis vectors $|g\rangle$ can
be zero-eigenvectors. However, we cannot in principle
exclude ``accidental'' zero-eigenvectors of the form of linear 
combinations
$\sum_i \alpha_i |g_i\rangle$. In case such vectors exist, we will
simply adopt the standard procedure of
defining our physical Hilbert space as the quotient space
$H^{(N)}_{ph}=H^{(N)}/{\cal N}^{(N)}$, where ${\cal N}^{(N)}$
denotes the span of all zero-eigenvectors.

Lastly, the boundedness of the transfer matrix (and therefore also of
$\hat T_{N}^{2}$) for finite spatial volumes
$N$ follows from the fact that (i) there is only a finite number of 
eigenvalues since the Hilbert space $H^{(N)}$ is finite-dimensional, 
and (ii) each matrix element $\langle g_{2}|\hat T_{N}|g_{1}\rangle$
has a finite value, because it has the form of a {\it finite} sum of
terms ${\rm e}^{-S}/C(T)$. (Note that this need not be
true in general if we abandoned the simplicial manifold restrictions, 
because
then the number of interpolating geometries for given, fixed boundary 
geometries would not necessarily be finite.)

\section{Monte Carlo moves}\label{monte}

In numerical simulations of the discrete Lorentzian model, as well as 
for some theoretical considerations, one needs 
so-called ``moves'', that is, a set of basic (and usually local)
manipulations of a triangulation, resulting in a different 
triangulation within the class under consideration. The relevant class of 
geometries for our purposes is that of Lorentzian (three- or 
four-dimensional) triangulations with a fixed number $t$ of 
time-slices, with finite space-time volume, and obeying simplicial 
manifold constraints. The set of basic moves should be ergodic, that 
is, one should be able to get from any triangulation to any other by
a repeated application of moves from the set. 

The Monte Carlo moves used in Euclidean dynamical triangulations
\cite{mcmoves,iceland} are not directly applicable since in general 
they do not respect
the sliced structure of the Lorentzian discrete geometries. 
Our strategy for constructing suitable sets of moves
is to first select moves that are ergodic {\it within}
the spatial slices $t=const$ (this is clearly a necessary condition for
ergodicity in the full triangulations), and then supplement them
by moves that act within the sandwiches $\Delta t\equ 1$. 
In $d\equ 3$, this results in a set of five basic moves 
which has already been used in numerical 
simulations \cite{ajl1,ajl2}, and which we believe is ergodic. 
For completeness, we will describe them again below. We will also
propose an analogous set of moves in four dimensions, which
are somewhat more difficult to visualize. 
In manipulating triangulations, especially in four dimensions,
a useful mental aid are the numbers of simplicial building blocks of 
type A contained in a given building block of type B of equal or 
larger dimension (Table 1).

As usual, if the moves are used as part of a Monte Carlo updating
algorithm, they will be rejected in instances when the resulting
triangulation would violate the simplicial manifold constraints.
The latter are restrictions on the possible gluings (the pair-wise 
identifications of sub-simplices of dimension $d-1$) to make
sure that the boundary of the neighbourhood of each vertex has
the topology of a $S^{d-1}$-sphere \cite{mauroetal}. In dimension 
three, ``forbidden'' configurations are conveniently
characterized in terms of the intersection pattern of the
triangulation at half-integer times (see \cite{ajl1}, Appendix 2). 
This is given 
by a two-dimensional tesselation in terms of squares and triangles or, 
alternatively, its dual graph with three- and four-valent crossings. 
Also in four dimensions, one may characterize sandwich geometries by 
the three-dimensional geometry obtained when cutting them at height
$1/2$. It is formed by two types of three-dimensional building blocks, 
a tetrahedron and a ``thickened'' triangle.

\begin{table}
	\begin{center}
	\renewcommand{\arraystretch}{1.25}
	\begin{tabular}{ |c||c|c|c|c|c|c|c|c|c|c|}
		\hline
		contains &
		$N_{0}$ & $N_{1}^{\rm TL}$ & $N_{1}^{\rm SL}$ & 
		$N_{2}^{\rm TL}$ & $N_{2}^{\rm SL}$ &
 	    $N_{3}^{(3,1)}$ & $N_{3}^{(2,2)}$ & $N_{3}^{\rm SL}$ & 
 	    $N_{4}^{(4,1)}$ & $N_{4}^{(3,2)}$ \\
		\hline\hline
		$N_{0}$ & 1 & 2 & 2 & 3 & 3 & 4 & 4 & 4 & 5 & 5  \\
		\hline
		$N_{1}^{\rm TL}$ &   & 1 & 0 & 2 & 0 & 3 & 4 & 0 & 4 & 6  \\
		\hline
		$N_{1}^{\rm SL}$ &   &   & 1 & 1 & 3 & 3 & 2 & 6 & 6 & 4  \\
		\hline
		$N_{2}^{\rm TL}$ &   &   &   & 1 & 0 & 3 & 4 & 0 & 6 & 9  \\
		\hline
		$N_{2}^{\rm SL}$ &   &   &   &   & 1 & 1 & 0 & 4 & 4 & 1  \\
		\hline
		$N_{3}^{(3,1)}$ &   &   &   &   &   & 1 & 0 & 0 & 4 & 2  \\
		\hline
		$N_{3}^{(2,2)}$ &   &   &   &   &   &   & 1 & 0 & 0 & 3  \\
		\hline
		$N_{3}^{\rm SL}$ &   &   &   &   &   &   &   & 1 & 1 & 0  \\
		\hline
		$N_{4}^{(4,1)}$ &   &   &   &   &   &   &   &   & 1 & 0  \\
		\hline
		$N_{4}^{(3,2)}$ &   &   &   &   &   &   &   &   &   & 1  \\
		\hline
	\end{tabular}
	\end{center}	
	\label{ntable}
	\caption{Numbers of simplicial building blocks contained in simplices 
	of equal or higher dimension.}
\end{table}

\subsection{Moves in three dimensions}

As already described in \cite{ajl1}, there are five basic moves
(counting inverse moves as separate). They map 
one 3d Lorentzian triangulation 
into another, while preserving the constant-time slice structure, as 
well as the total proper time $t$. 
We label the moves by how they affect the number of simplices of
top-dimension, i.e. $d\equ 3$. 
A $(m,n)$-move is one that operates on
a local sub-complex of $m$ tetrahedra and replaces it by a different
one with $n$ tetrahedra. The tetrahedra themselves are characterized
by 4-tuples of vertex labels. Throughout, we will not distinguish 
moves that are mirror images of each other under time reflection.
In detail, the moves are 
\begin{itemize}
\item[(2,6):] 
this move operates on a pair of a (1,3)- and a (3,1)-tetrahedron 
(with vertex labels 1345 and 2345) sharing
a spatial triangle (with vertex labels 345). A vertex (with label 6)
is then inserted at the centre of the triangle and connected by new
edges to the vertices 1, 2, 3, 4 and 5. The final configuration
consists of six tetrahedra, three below and three above the spatial
slice containing the triangles (Fig.\ \ref{26m}). 
This operation may be encoded by writing
\begin{equation}
1\underline{345} +2\underline{345} \rightarrow 
1\underline{346} +2\underline{346} +
1\underline{356} +2\underline{356} +
1\underline{456} +2\underline{456},
\label{26move}
\end{equation}
where the underlines indicate simplices shared by several tetrahedra
(not necessarily all, in order not to clutter the notation).
The inverse move (6,2) corresponds to a reversal of the arrow in
(\ref{26move}). Obviously, it can only be performed if 
the triangulation contains a suitable sub-complex of six tetrahedra.
\begin{figure}[t]
\centerline{\scalebox{0.6}{\rotatebox{0}{\includegraphics{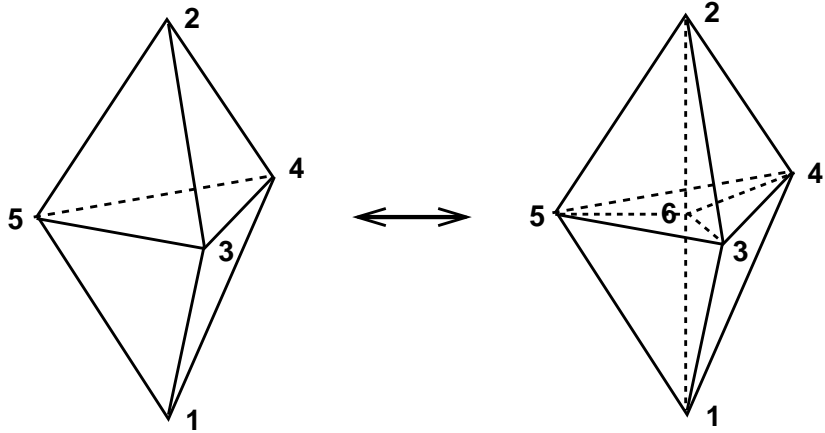}}}}
\caption[26m]{The (2,6)-move in three dimensions.}
\label{26m}
\end{figure}
\item[(4,4):] 
This move can be performed on a sub-complex of two (1,3)- and
two (3,1)-tetrahedra
forming a ``diamond'' (see Fig.\ \ref{44m}),
with one neighbouring pair each above and below a spatial slice.
The move is then 
\begin{equation}
1\underline{235}+\underline{235}6+1\underline{345}+\underline{345}6
\rightarrow 1\underline{234} +\underline{234}6+1\underline{245}
+\underline{245}6.
\label{44move}
\end{equation}
From the point of view of the spatial ``square'' (double triangle) 2345,
the move (\ref{44move}) corresponds to a flip of its diagonal.
It is accompanied by a corresponding
reassignment of the tetrahedra constituting
the diamond. 
\begin{figure}[t]
\centerline{\scalebox{0.6}{\rotatebox{0}{\includegraphics{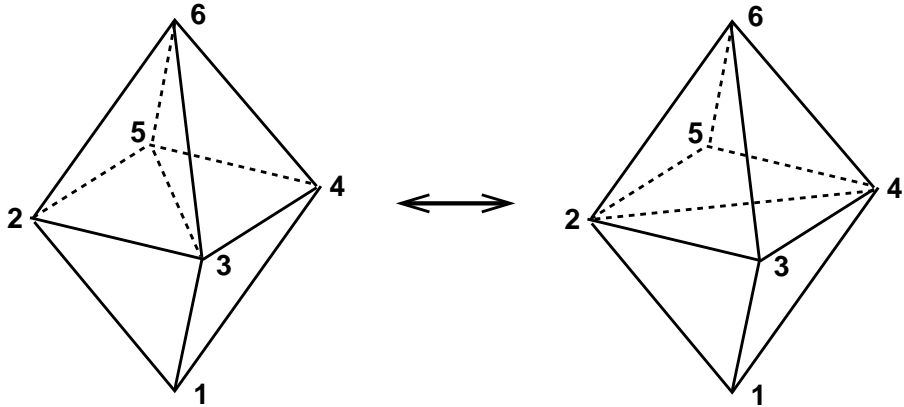}}}}
\caption[44m]{The (4,4)-move in three dimensions.}
\label{44m}
\end{figure}
The (2,6)- and (6,2)-moves, together with the (4,4)-move (which is its 
own inverse) induce moves within the spatial slices that are known
to be ergodic for two-dimensional triangulations.
\item[(2,3):] 
The last move, together with its inverse, affects the sandwich 
geometry without changing the spatial slices at integer-$t$. 
It is performed on a pair of a (3,1)- and a (2,2)-tetrahedron which 
share a triangle 345 in common (see Fig.\ \ref{23m}), 
and consists in substituting this 
triangle by the one-dimensional edge 12 dual to it,
\begin{equation}
1\underline{345}+2\underline{345}\rightarrow 
\underline{12}34+\underline{12}35+\underline{12}45.
\label{23move}
\end{equation}
The resulting configuration consists of one (3,1)- and two 
(2,2)-tetrahedra, sharing the link 12. Again, there is an obvious 
inverse.
\begin{figure}[t]
\centerline{\scalebox{0.6}{\rotatebox{0}{\includegraphics{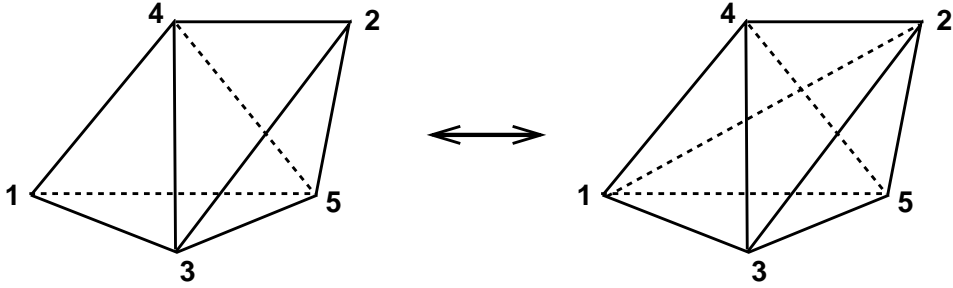}}}}
\caption[23m]{The (2,3)-move in three dimensions.}
\label{23m}
\end{figure}
\end{itemize}

\subsection{Moves in four dimensions}

If we distinguish between the space- and time-like character of
all of the four-dimensional moves, there is a total of ten moves
(including inverses). We will again characterize simplices in terms of 
their vertex labels. The first two types of moves, (2,8) and (4,6),
reproduce a set of ergodic moves in three dimensions when restricted 
to spatial slices. We will describe each of the moves in turn.
\begin{itemize}
\item[(2,8):] The initial configuration for this move is a pair of
a (1,4)- and a (4,1)-simplex, 
sharing a purely space-like tetrahedron 3456 (Fig.\ \ref{28m}).
\begin{figure}[t]
\centerline{\scalebox{0.6}{\rotatebox{0}
{\includegraphics{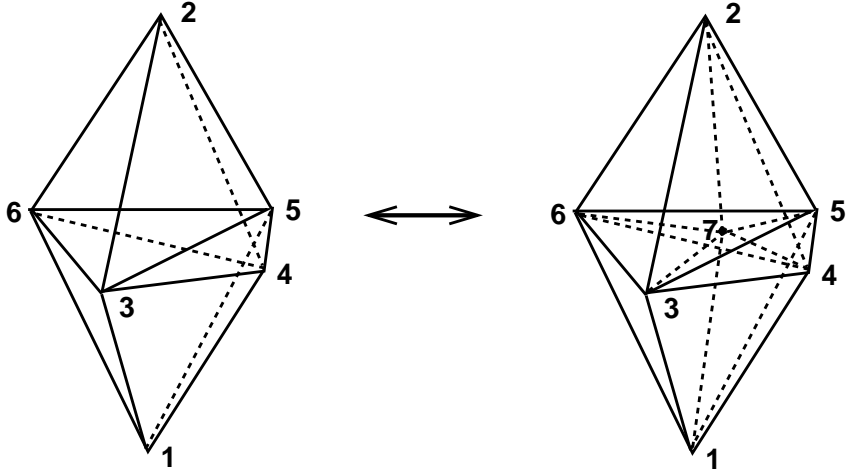}}}}
\caption[28m]{The (2,8)-move in four dimensions.}
\label{28m}
\end{figure}
The move consists in inserting an additional vertex 7 at the centre 
of this tetrahedron and subdiving the entire sub-complex so as to 
obtain eight four-simplices,
\begin{equation}
1\underline{3456}+2\underline{3456}\rightarrow
1\underline{3457}+2\underline{3457}+
1\underline{3467}+2\underline{3467}+
1\underline{3567}+2\underline{3567}+
1\underline{4567}+2\underline{4567},
\label{28move}
\end{equation}
with an obvious inverse.
\item[(4,6):] In this configuration, we start from a pair (2345,3456) 
of spatial tetrahedra sharing a common triangle 345, which are 
connected to a vertex 1 at time $t\mi 1$ and another vertex 7 at time
$t\pl 1$, forming together a sub-complex of size six (Fig.\ \ref{46m}). 
The move
consists in swapping the triangle 345 with its (spatially) dual edge 26,
during which the two spatial tetrahedra are substituted by three,
and the geometry above and below the spatial slice at time $t$
changed accordingly. In our by now familiar notation, this amounts to
\begin{equation}
1\underline{2345} +\underline{2345}7 +
1\underline{3456} +\underline{3456}7 \rightarrow
1\underline{2346} +\underline{2346}7 +
1\underline{2356} +\underline{2356}7 +
1\underline{2456} +\underline{2456}7,
\label{46move}
\end{equation}
with the arrow reversed for the corresponding inverse move. 
\begin{figure}[t]
\centerline{\scalebox{0.6}{\rotatebox{0}
{\includegraphics{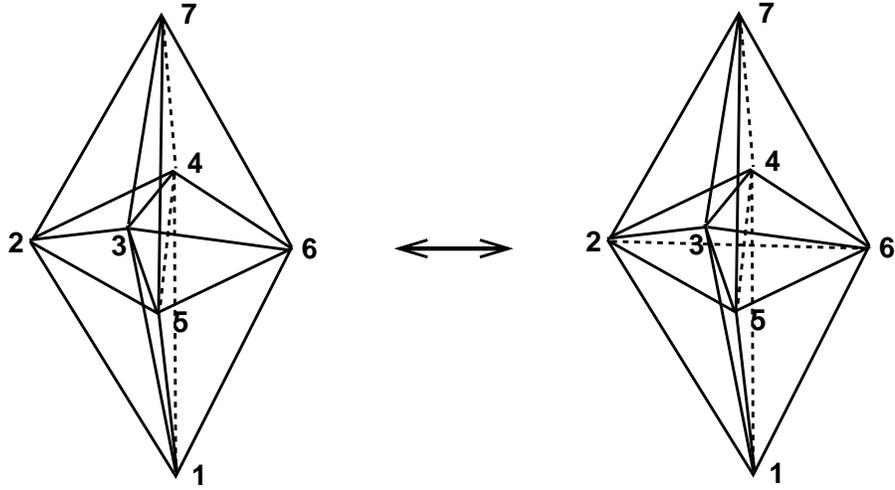}}}}
\caption[46m]{The (4,6)-move in four dimensions.}
\label{46m}
\end{figure}
\item[(2,4):] This type of move comes in two varieties. 
Its general structure is as follows:
the initial configuration is a pair of 4-simplices with
a common tetrahedron. During the move, this tetrahedron
is ``deleted'' and substituted by its (in a four-dimensional
sense) dual edge. The end result is a sub-complex consisting of four 
4-simplices. From a Lorentzian point of view, there are two
situations where the application of this move does not interfere with 
the slice-structure or the manifold constraints. In the first one, 
a (4,1)- and a (3,2)-tetrahedron from the same sandwich share 
a (3,1)-tetrahedron 3456 (Fig.\ \ref{24ma}). 
The dual edge 12 is time-like and
shared in the final configuration by one (4,1)- and three 
(3,2)-simplices. The second possibility is that of two
(3,2)-simplices sharing a (2,2)-tetrahedron. One of the 
(3,2)-simplices is ``upside-down'', such that the entire sub-complex
has spatial triangles in both the slices at $t$ and at $t\pl 1$
(134 and 256, see Fig.\ \ref{24mb}). After the move, the total number of
(3,2)-simplices has again increased by two. (Note that the
changes of the bulk variables are encoded in the $f$-vectors of
Sec.\ \ref{kinema}.) Both types of moves are described by the relation
\begin{equation}
1\underline{3456}+2\underline{3456}\rightarrow \underline{12}345 +
\underline{12}346 +\underline{12}456 +\underline{12}356,
\label{24move}
\end{equation}
and their inverses by the converse relation.
\begin{figure}[t]
\centerline{\scalebox{0.6}{\rotatebox{0}
{\includegraphics{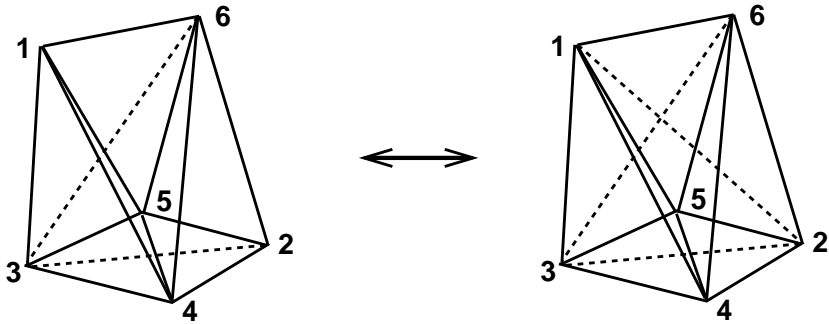}}}}
\caption[24ma]{The (2,4)-move in four dimensions, first version.}
\label{24ma}
\end{figure}
\begin{figure}[t]
\centerline{\scalebox{0.6}{\rotatebox{0}
{\includegraphics{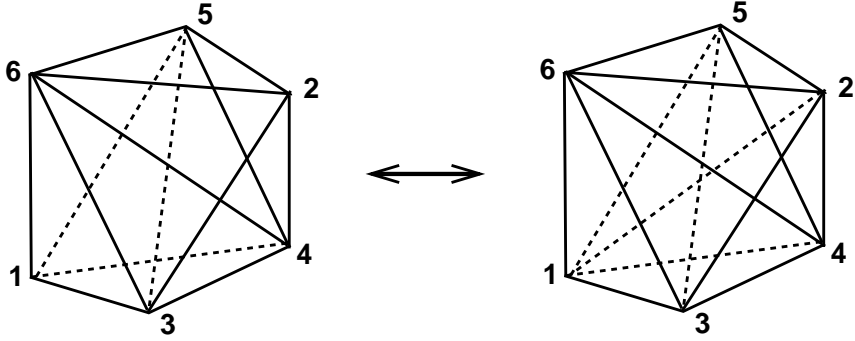}}}}
\caption[24mb]{The (2,4)-move in four dimensions, second version.}
\label{24mb}
\end{figure}
\item[(3,3):] The initial sub-complex in this type of move is made up 
of three 4-simplices which share a triangle in common. In the course 
of the move, this triangle is ``deleted'' and substituted by its
dual (in a four-dimensional sense), which is again a triangle. 
It is straightforward to verify that this move can only occur
in Lorentzian gravity if both of the triangles involved are time-like.
Again, there are two allowed variations of the move. In the first 
one, both the initial and final sub-complex consist of one (4,1)- and 
two (3,2)-simplices, and the space-like edge of each of the triangles 
123 and 456 lies in the same slice $t\equ const$ (Fig.\ \ref{33ma}). 
The 4-simplices are rearranged according to
\begin{equation}
12\underline{456}+13\underline{456}+23\underline{456}
\rightarrow \underline{123}45+\underline{123}46+\underline{123}56.
\label{33move}
\end{equation}
\begin{figure}[t]
\centerline{\scalebox{0.6}{\rotatebox{0}
{\includegraphics{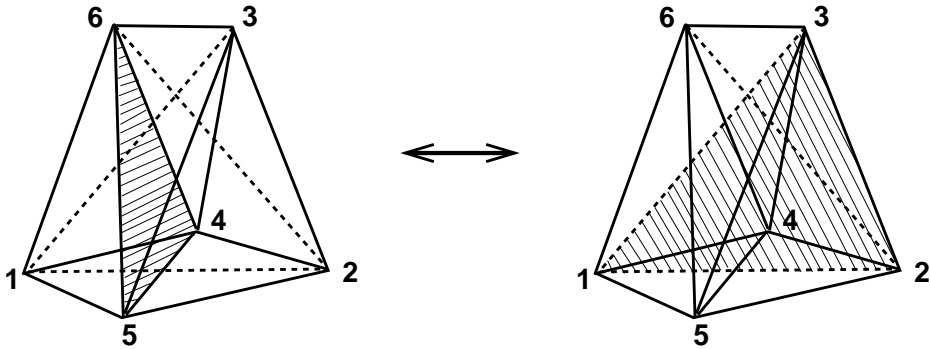}}}}
\caption[33ma]{The (3,3)-move in four dimensions, first version.}
\label{33ma}
\end{figure}
\begin{figure}[t]
\centerline{\scalebox{0.6}{\rotatebox{0}
{\includegraphics{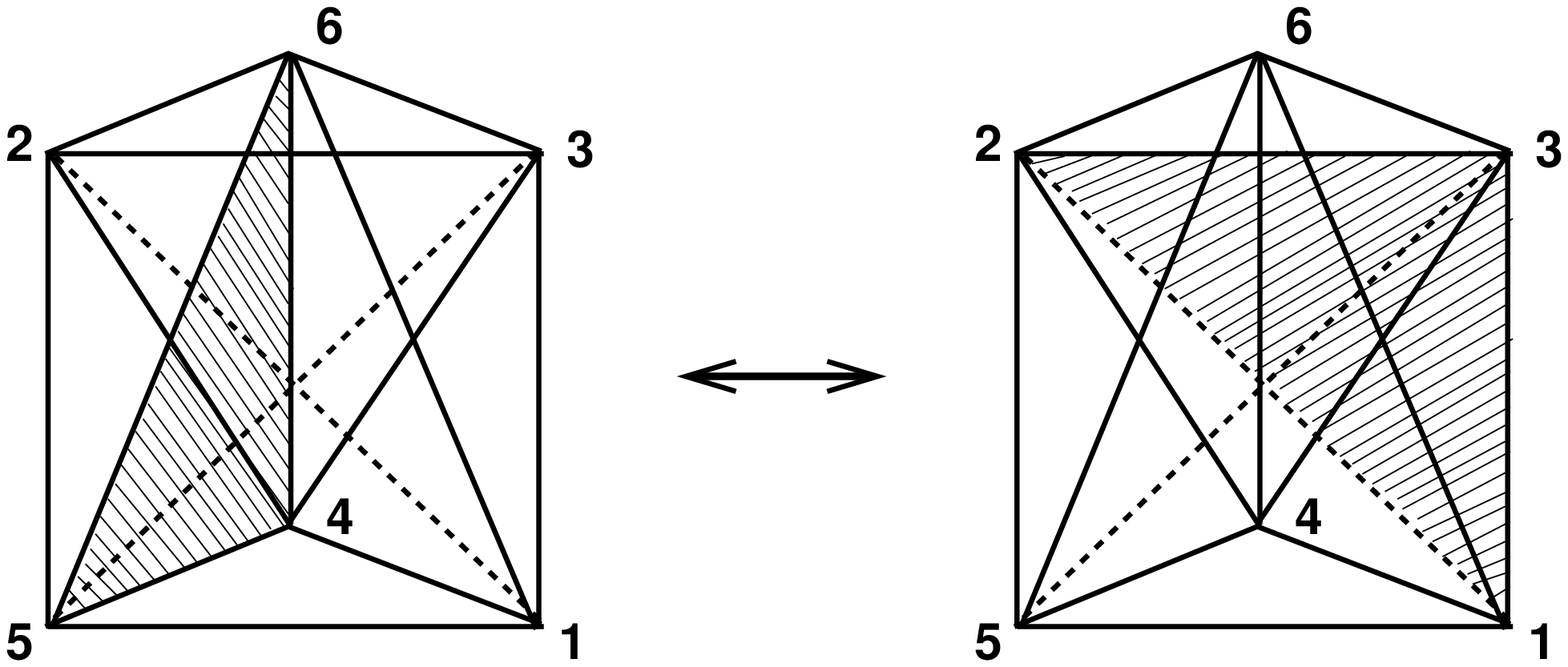}}}}
\caption[33mb]{The (3,3)-move in four dimensions, second version.}
\label{33mb}
\end{figure}
The other initial configuration for which (\ref{33move}) can be
performed involves only (3,2)-simplices of both orientations, as 
illustrated in Fig.\ \ref{33mb}. 
The ``swapped'' triangle 123 now has its space-like
edge in the opposite spatial slice from that of the original triangle 
456. As in the (4,4)-move in three dimensions, this type of move 
is its own inverse.

\end{itemize}

\section{Kinematic bounds}\label{kinema}

Without solving the combinatorics of the Lorentzian gravity models 
explicitly, some qualitative insights can be gained by studying
the behaviour of their partition functions for either very small or 
very large values of the inverse Newton's constant $k$. 
By comparing our results with those of {\it Euclidean} dynamical 
triangulations in 3 and 4d \cite{gab,crump}, we will see that 
significant differences arise already at this kinematic level. 
Since the partition function for Lorentzian gravity is evaluated after 
Wick-rotating, we will in the following set $\alpha\equ -1$, and
work in the Euclidean sector of our models. We will perform the 
analysis for $d=3$ and $d=4$ in turn.

\subsection{Analysis in three dimensions}

For ease of comparison with the Euclidean results, we write the
Regge action in standard form (c.f. (\ref{3dactfinal})), namely, 
as
\beq{euclact1}
S_{\rm EDT}^{(3)}=k_3 N_3 -k_1 N_1,
\eeq
with the couplings
\beq{coupl1}
k_1=2\pi k,\qquad
k_3=6\  k \arccos\frac{1}{3}+
\lambda \frac{1}{6\sqrt{2}}. 
\eeq
Lorentzian triangulations $T$ are conveniently characterized by
their $f$-vectors $f(T)$, which keep track of the
numbers (\ref{nnumbers}) of simplicial building blocks. 
In three dimensions, we have
\begin{equation}
f^{(3)}=(N_{0},N_{1}^{\rm SL}, N_{1}^{\rm TL}, N_{2}^{\rm SL}, 
N_{2}^{\rm TL}, N_{3}^{(3,1)},N_{3}^{(2,2)}).
\label{fvec3}
\end{equation}
Rewriting the action as
\begin{equation}
S_{\rm EDT}^{(3)}=N_{3} (k_3 -k_1 \frac{N_1}{N_{3}})=:
N_{3} (k_3 -k_1 \xi),
\end{equation}
we see that for large coupling $k_{1}$ and at fixed space-time volume,
configurations with a large value of $\xi$ will be energetically 
preferred. In the Euclidean case, it can be shown that in the
thermodynamic limit $N_{3}\rightarrow\infty$, the ratio $\xi$ obeys
the kinematic bounds
\begin{equation}
1\leq \xi\leq \frac{4}{3},\;\;\;\;\;\mbox{Euclidean 3d triangulations}.
\label{kineu3}
\end{equation}
The upper bound is saturated for so-called branched-polymer
(or ``stacked-sphere'') configurations. Geometrically 
degenerate configurations of this type dominate the partition 
function in the 
continuum limit and seem to be responsible for the failure of the
{\it Euclidean} model to lead to an interesting continuum theory, 
both in three and four dimensions. 

Obviously, the geometries of the Lorentzian model must satisfy 
(\ref{kineu3}), which is valid for all 3d simplicial manifolds.
However, it turns out that the Lorentzian dynamical triangulations 
satisfy a more stringent upper bound. Using identities from Sec.\ 
\ref{topo}, we can write
\begin{equation}
N_{3}=N_{3}^{(2,2)}+N_{3}^{(3,1)}=N_{3}^{(2,2)}+4 N_{0} -4\chi t
\geq N_{3,{\rm min}}^{(2,2)}+4 N_{0} -4\chi t.
\label{n3count}
\end{equation}
The minimal $N_{3}^{(2,2)}$ is given by 
$N_{3,{\rm min}}^{(2,2)} =c_{\rm min}t$, where $c_{\rm min}>0$ is
an integer counting the minimal possible number of
(2,2)-tetrahedra per time-step $\Delta t\equ 1$. Using further 
identities, (\ref{n3count}) therefore becomes
\begin{equation}
4 N_{1}\leq 5 N_{3}-(c_{\rm min}-4\chi ) t.
\end{equation} 
From this we deduce that in the large-volume limit (where
$t/N_{3}\rightarrow 0$) the upper bound for $\xi$ is lowered to $5/4$, 
so that the Euclidean relation (\ref{kineu3}) is
substituted by
\begin{equation}
1\leq \xi\leq \frac{5}{4},\;\;\;\;\;\mbox{Lorentzian 3d triangulations}.
\label{kinlor3}
\end{equation}
We conclude that the branched-polymer regime of
Euclidean gravity cannot be realized in the Lorentzian case.
The upper bound in (\ref{kinlor3}) is saturated when the number
of (2,2)-simplices is minimal. 

There is an alternative way of deriving (\ref{kinlor3}), which also
sheds some light on the nature of the lower bound.
Here, one asks whether the bounds on $\xi$ can be 
approached by a successive application of the Monte Carlo moves 
described in Sec.\ \ref{monte}. 
In three dimensions, the effect of the Monte Carlo 
moves on the $f$-vector is
\begin{eqnarray}
&&\Delta_{(2,6)} f^{(3)}=(1,3,2,2,6,4,0)\label{delf31},\\
&&\Delta_{(4,4)} f^{(3)}=(0,0,0,0,0,0,0)\label{delf32},\\
&&\Delta_{(2,3)} f^{(3)}=(0,0,1,0,2,0,1)\label{delf33},
\end{eqnarray}
with obvious expressions for the inverse moves.
Let us now choose some initial configuration $f_{0}^{(3)}$, for example, 
a $t$-translation-invariant geometry of minimal volume of the
form $f_{0}^{(3)}=\vec c\ t$, where $\vec c$ is a constant 
seven-vector characterizing a sandwich geometry. Any
triangulation that is reached from the initial one by a sequence of
allowed Monte Carlo moves has an $f$-vector
\begin{equation}
f^{(3)}=(c_{1}t+x, c_{2}t+3 x, c_{3}t+2 x+y,  
c_{4}t +2 x, c_{5}t+6 x+2 y, c_{6}t+4 x, c_{7}t+y ).
\label{fullf3d}
\end{equation}
The integers $x=n_{(2,6)}-n_{(6,2)}$ and $y=n_{(2,3)}-n_{(3,2)}$ 
count the ``excess'' of applications of a move over its inverse. 
From (\ref{fullf3d}) we can read off the value of $\xi$ as
\begin{equation}
\xi =\frac{N_{1}}{N_{3}} =\frac{(c_{2}+c_{3})t+5 x+y}{(c_{6}+c_{7})t +
4 x+y}.
\label{xi3d}
\end{equation}
It is obvious that for large triangulations this 
ratio is maximized by applying only (2,6)-moves. This is always
possible, since this Monte Carlo move is a subdivision move that can 
always be performed. In the infinite-volume limit 
(this also implies $x \gg t$), 
$\xi$ approaches $5/4$, in 
accordance with our earlier result. Alternatively, a triangulation
can be enlarged by applying (2,3)-moves, which according to 
(\ref{delf33}) increase the
number of (2,2)-tetrahedra. 
It is not immediately clear that we can saturate the
lower bound in (\ref{kinlor3}) by taking $y\rightarrow\infty$
since $y$ is not entirely independent of $x$ 
(the applicability of the (2,3)-move is related to the availability
of (3,1)-tetrahedra, which in turn is governed by $x$). 
Nevertheless, in the large-volume
limit such configurations do actually exist, as 
we will demonstrate below.

Another type of degenerate geometry that was observed to dominate the
state sums of Euclidean dynamical triangulations at {\it small}
$k$ \cite{hottaetal,catteralletc} is characterized by an inequality
which is non-linear in the bulk variables, namely,
\beq{crumpled}
N_1\leq \frac{1}{2} N_0 (N_0 -1). 
\eeq
A triangulation saturating (\ref{crumpled}) is called
``2-neighbourly'' \cite{crump} since any of its vertices is 
connected by an edge to any other vertex. The resulting space
is of maximal Hausdorff dimension, because ``one can get anywhere in a 
single step''. Such singular structures have indeed been observed
in the so-called crumpled phase of Euclidean quantum gravity in three 
and four dimensions. 

Again, it is impossible to come anywhere near this phase in
the continuum limit of the Lorentzian model. This has to do with the
fact that by construction of the Lorentzian geometries, two vertices 
can only be connected by a link if they lie in the same or 
in successive spatial slices. Instead of 
(\ref{crumpled}), we therefore have separate relations for the numbers 
$N_1^{{\rm SL}}$ and $N_1^{{\rm TL}}$ of space- and time-like
edges,
\beq{inequa}
N_1^{\rm SL}=\sum_{\tau =0}^{t-1} (3 N_0(\tau)-6)=3 N_0-6 t,\qquad
N_1^{\rm TL}\leq \sum_{\tau =0}^{t-1} N_0(\tau) N_0(\tau +1).
\eeq
Assuming canonical scaling, the right-hand side of the
inequality (\ref{crumpled}) behaves like (length)$^6$, whereas the
second relation in (\ref{inequa}) scales only like
(length)$^5$, proving our earlier statement. 

\begin{figure}[t]
\centerline{\scalebox{0.6}{\rotatebox{0}{\includegraphics{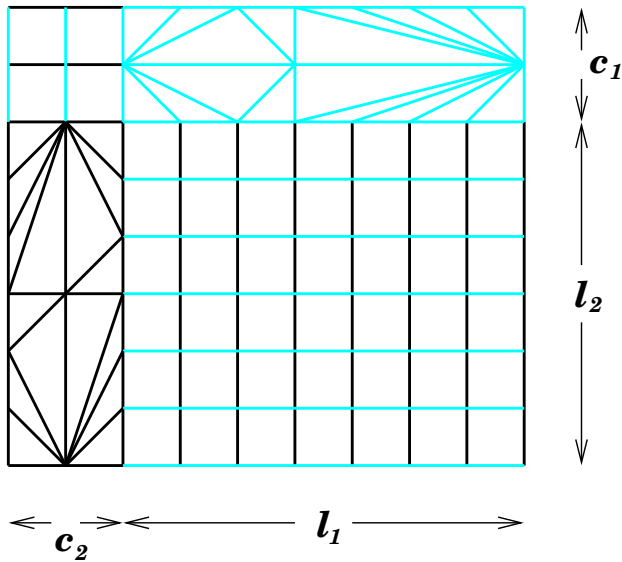}}}}
\caption[crumple]{One of a sequence of toroidal sandwich configurations 
at $\tau +1/2$ which saturate the inequality in (\ref{inequa})
in the large-volume limit (opposite sides to be identified). 
Light links correspond to triangle edges at time $\tau$, dark links
to triangle edges at time $\tau +1$.}
\label{crumple}
\end{figure}
The inequality in (\ref{inequa}) can indeed be saturated by 
certain regular Lorentzian geometries in the large-volume limit, as 
one can show by explicit construction. For ease of 
visualization, let us assume that the two-dimensional slices have
the topology of two-tori $T^{2}$. Consider a sandwich
where the initial two-geometry at time $\tau$ is a ``strip'' of
discrete length $l_{1}$ and width $c_1=2$,
consisting of $2 l_{1}+4$ triangles\footnote{The +4 comes from the
fact that the minimal length of the intermediate layer of the
two-torus must be $l=2$, to give a simplicial 2d manifold under gluing.}, 
with opposite sides
identified to give a torus. Similarly, at time $\tau +1$, take
a strip of length $l_{2}$ and width $c_{2}=2$ with
$2 l_{2}+4$ triangles. Choose the 
interpolating geometry between the two tori such that the
(2,2)-tetrahedra when viewed at time $\tau +1/2$
form two rectangular blocks, of size $l_{1} l_{2}$ and of
size $c_{1} c_{2}$ (Fig.\ \ref{crumple}). 
Consider now a sequence of such configurations
where both of the $l_{i}$ become large, but
the integers $c_{i}$ remain unchanged. Because of $N_{0}(\tau)=
N_{1}^{\rm SL}(\tau)/3 =l_{1}+2$ and
$N_{0}(\tau +1)=N_{1}^{\rm SL}(\tau +1)/3 =l_{2}+2$, we find
that in the large-volume limit the bulk variables associated
with the sandwich are given by
$N_{1}\approx N_{1}^{\rm TL}\approx l_{1} l_{2}$, and by
$N_3 =N_2^{\rm SL}(\tau)+N_2^{\rm SL}(\tau +1)+ N_3^{(2,2)}\approx
l_1 l_2$. For the vertex numbers, we have
$N_0(\tau)N_0(\tau +1)\approx l_1 l_2$. This verifies our earlier
claims that both the inequality (\ref{inequa}) and the lower bound
in (\ref{kinlor3}) are saturated by this class of configurations.
Of course, we achieved this by a ``fine-tuning'' of geometry,
and the configurations thus obtained are highly
non-generic. 

The above considerations show that the phase structure of 
three-dimensional Lorentzian 
gravity must be quite different from that of the Euclidean theory
since already at a kinematic level, the extreme branched-polymer and 
crumpled configurations cannot be realized. Obviously it does not 
exclude the possibility that milder pathologies prevail in the
continuum limit, but it is clearly a step in the right direction.
Hopes that the Lorentzian model is therefore
better-behaved than its Euclidean counterpart have been boosted
by recent numerical investigations in three dimensions 
\cite{ajl1,ajl2}. There it was found that for a large range of the 
gravitational coupling constant $k_{1}$, the partition function is 
dominated by a phase of extended geometry (and the parameter $\xi$
stays well away from its extremal bounds (\ref{kinlor3})), 
whose global scaling 
properties are those of a three-dimensional object. Compared with the 
Euclidean case, this is a completely new and very promising phenomenon.
It reiterates the finding in $d=2$ \cite{al} that the causal structure 
of the Lorentzian space-times acts effectively as a regulator for the 
quantum geometry.

\subsection{Analysis in four dimensions}

The analysis of the kinematic bounds in four dimensions
is completely analogous to that of the previous subsection,
and we will refrain from spelling out all the details.
We start from the action in the form 
(c.f. (\ref{4dactfinal}))
\beq{euclact2}
S_{\rm EDT}^{(4)}=k_4 N_4 -k_2 N_2
=N_{4} (k_4 -k_2 \frac{N_2}{N_{4}})=:
N_{4} (k_4 -k_2 \xi'),
\label{actn4}
\eeq
with the coupling constants
\beq{coupl2}
k_2=  \frac{\sqrt{3}}{2}\pi k,\qquad
k_4= \frac{5 \sqrt{3}}{2}  k \arccos \frac{1}{4} +
\lambda \frac{\sqrt{5}}{96}.  
\eeq
In (\ref{actn4}) we have already anticipated that a convenient order 
parameter in four dimensions is the quotient $\xi'=N_{2}/N_{4}$.
The $f$-vector has now length 10, and is given by
\begin{equation}
f^{(4)}=(N_{0},N_{1}^{\rm SL}, N_{1}^{\rm TL}, N_{2}^{\rm SL}, 
N_{2}^{\rm TL},N_{3}^{\rm SL}, N_{3}^{(3,1)},N_{3}^{(2,2)},
N_{4}^{(4,1)},N_{4}^{(3,2)}).
\label{fvec4}
\end{equation}
The discussion parallels closely the three-dimensional case, with 
the relevant Euclidean inequalities \cite{gab,crump}
\begin{equation}
2\leq \xi'\leq \frac{5}{2},\;\;\;\;\;\mbox{Euclidean 4d triangulations}.
\label{kineu4}
\end{equation}
Let us derive their Lorentzian analogues by analyzing the effect of the 
Monte Carlo moves of Sec.\ \ref{monte} on the $f$-vector (\ref{fvec4}), 
which is
\begin{eqnarray}
&&\Delta_{(2,8)} f^{(4)}=(1,4,2,6,8,3,12,0,6,0)\label{delf41},\\
&&\Delta_{(4,6)} f^{(4)}=(0,1,0,2,2,1,4,0,2,0)\label{delf42},\\
&&\Delta_{(2,4)} f^{(4)}=(0,0,1,0,4,0,2,3,0,2)\label{delf43},\\
&&\Delta_{(3,3)} f^{(4)}=(0,0,0,0,0,0,0,0,0,0)\label{delf44}.
\end{eqnarray}
Starting from a minimal initial configuration $f_{0}^{(4)}$, and
defining $x=n_{(2,8)}-n_{(8,2)}$, $y=n_{(4,6)}-n_{(6,4)}$ and
$z=n_{(2,4)}-n_{(4,2)}$, and dropping terms proportional to $t$ in the 
large-volume limit, one obtains
\begin{equation}
\xi' =\frac{N_{2}}{N_{4}}\longrightarrow\frac{7 x+2 (y+z)}{3 x+ y+z}.
\label{xi4d}
\end{equation}
Again, $\xi'$ is maximized by applying the subdivision move (2,8)
only. Note also that -- like in three dimensions -- this corresponds
to creating stacked-sphere configurations within the spatial
slices of constant integer-$t$. (We are using the expression
``stacked spheres'' in a loose sense, and only require that the
triangulations look like stacked spheres almost everywhere.)
For the purposes of relation
(\ref{xi4d}), the moves (4,6) and (2,4) play an equivalent role,
although they generate different types of four-simplices. It is
clear that a lower bound for the ratio is given by 2, although
again it is not immediately clear that this value can actually be
attained by any triangulation in the thermodynamic limit.
Like in 3d, we can conclude that the range for the parameter
$\xi'$ in Lorentzian gravity is reduced, such that
\begin{equation}
2\leq \xi'\leq \frac{7}{3},\;\;\;\;\;\mbox{Lorentzian 4d triangulations}.
\label{kinlor4}
\end{equation}
We leave it as an exercise to the reader to prove this relation
directly from (\ref{kineu3}) and the identities in Sec.\ 3, 
as well as to construct explicit triangulations that saturate
the lower bound.
Again, our analysis implies the absence of the polymeric geometries
found in the Euclidean theory. (Note that a polymeric behaviour
is in principle allowed both in three- and four-dimensional
Lorentzian triangulations, as long as it takes place
{\it in the spatial directions only}. In fact, this corresponds
precisely to the limit $x\rightarrow\infty$ in both (\ref{xi3d})
and (\ref{xi4d}).) Likewise, there is no crumpled
phase. The argument of $d=3$ carries over to four dimensions,
with the only difference that the right-hand side of 
(\ref{crumpled}) scales like (length)$^8$, whereas the
corresponding relation in (\ref{inequa}) scales like
(length)$^7$.

\section{Summary and outlook}\label{summ}

In this paper, we have presented a comprehensive analysis of
a regularized path integral for gravity in terms of Lorentzian
dynamical triangulations. We have shown that for a finite
space-time volume $N$ the model is well-defined as a statistical 
system of geometry, with a finite state sum. 

Having established these properties, our main interest is the
behaviour of the system in a suitable continuum limit. 
For this, we must take the lattice volume $N\rightarrow\infty$
and the invariant lattice spacing $a\rightarrow 0$, and allow
for a renormalization of both the bare coupling constants and
the propagator itself. Although the discretized set-up of our model
looks rather similar in different dimensions, we expect
the details of the renormalization and the resulting
physics to depend strongly on $d$.

The continuum theory in $d=2$ is already explicitly known, since the
Lorentzian model has been solved exactly \cite{al,alnr}. It is obtained
by simply taking $N\rightarrow\infty$ or, equivalently, 
tuning the bare cosmological constant $\lambda$ to its critical
value. The theory describes a one-dimensional spatial ``quantum
universe'' whose volume fluctuates in time. Obviously the
physical significance of this model is somewhat obscure,
since there is no classical theory of 2d Einstein gravity,
and therefore no well-defined classical limit. Nevertheless,
it provides a beautiful example of the inequivalence of path
integrals of Euclidean and Lorentzian geometries. What is more,
we understand in a very explicit manner \cite{al,ackl} 
that the origin of the 
difference lies in the presence and absence of ``baby universes''
branching off in the time-direction (which in the Lorentzian case are 
incompatible with the ``micro-causality'' of the path-integral
histories).

``Solving the path integral'' in our formulation amounts to solving
the combinatorics of the Lorentzian triangulations in a given
dimension and for given boundary data, at least asymptotically.
Although this counting problem is mathematically well-defined, 
it is not simple when $d\geq 3$. In three space-time dimensions,
where the problem is presently under investigation,
an interesting relation with an already known matrix model with
ABAB-interaction has been uncovered \cite{ajlv}. 

A potentially powerful feature of our models
is the fact that they can also be studied numerically. Using Monte Carlo
methods, it was shown that 2d Lorentzian gravity coupled to matter
has no barrier at $c=1$ \cite{aal}, and that pure 3d Lorentzian gravity
possesses a ground state of extended geometry, with the global scaling 
properties of a three-dimensional space-time universe \cite{ajl1,ajl2}.
Both of these findings are highly significant, because they
show that these models lie in different universality classes from
their Euclidean counterparts. 

Let us recapitulate how this difference with the purely Euclidean
formulation came about. Our physical premise was to take the space of
{\it Lorentzian} space-times as our starting point. This enabled us
to impose certain causality constraints on the individual histories 
contributing to the path integral. As a consequence, 
like in classical relativity, ``time''- and ``space''-directions are 
{\it not} interchangeable, a feature that persists in the 
continuum quantum theory. In order for this philosophy to
lead anywhere (i.e. to yield convergent mathematical expressions) we
crucially needed a Wick rotation on the space of Lorentzian geometries.
The fact that such a map exists in our formulation is by no means
trivial. For example, if a similar construction were attempted
in Regge calculus, one would immediately run into trouble with
triangular inequalities.

Being familiar with the pitfalls of Euclidean gravitational path 
integrals, one may now wonder whether our path integrals after 
the Wick rotation to Euclidean signature do not also suffer from
pathologies, in the limit as $N\rightarrow\infty$.  
In particular, one may worry about the presence of a conformal 
divergence in $d\geq 3$ associated with the unboundedness of the 
gravitational
action. Remarkably, although such a divergence is {\it potentially}
present in our formulation, since configurations with large
negative action exist, these are entropically suppressed in
the continuum limit, at least in $d=3$ \cite{ajl1}. It has been
demonstrated in \cite{dl} how this phenomenon can be understood from 
the point of view of the {\it continuum} gravitational path integral:
a careful analysis of various Faddeev-Popov determinants 
reveals how the resulting path-integral {\it measure} can lead to
a non-perturbative cancellation of the conformal 
divergence\footnote{Note that the use of proper-time gauge in 
\cite{dl} made it
possible to define a non-perturbative Wick rotation analogous to
the one used in the present work.}.

To summarize, we have constructed a new and powerful
calculational tool for quantum gravity that in the best of all
worlds will lead
to a non-perturbative definition of an interacting quantum theory
of gravity in four dimensions. An inclusion of other matter
degrees of freedom is in principle straightforward.
At the discretized level,
our models are as well-defined as one could hope for.
An investigation in dimension 3, both analytically and numerically,
is well under way. It remains a major challenge to set up
sufficiently large numerical simulations in four dimensions. 
Clearly, because of the anisotropy between time and spatial
directions, larger lattice sizes than in the Euclidean dynamically
triangulated models will be needed. On the analytic front,
tackling the full combinatorics in four dimensions will
be difficult, as one would expect from any fully-fledged formulation
of quantum gravity. In order to approach the problem, we are
currently studying both the three-dimensional theory and its
relation to other formulations of 3d quantum gravity, as well as
gravitational models with extra symmetries.

\subsection*{Acknowledgements} All authors
acknowledge support by the
EU network on ``Discrete Random Geometry'', grant HPRN-CT-1999-00161, 
and by ESF network no.82 on ``Geometry and Disorder''.
In addition, J.A. and J.J. were supported by ``MaPhySto'', 
the Center of Mathematical Physics and Stochastics, financed by the 
National Danish Research Foundation, and
J.J. by KBN grants 2P03B 019\,17 and 998\,14.

\vfill\eject

\section*{Appendix 1: Lorentzian angles}

\begin{figure}[t]
\centerline{\scalebox{0.6}{\rotatebox{0}{\includegraphics{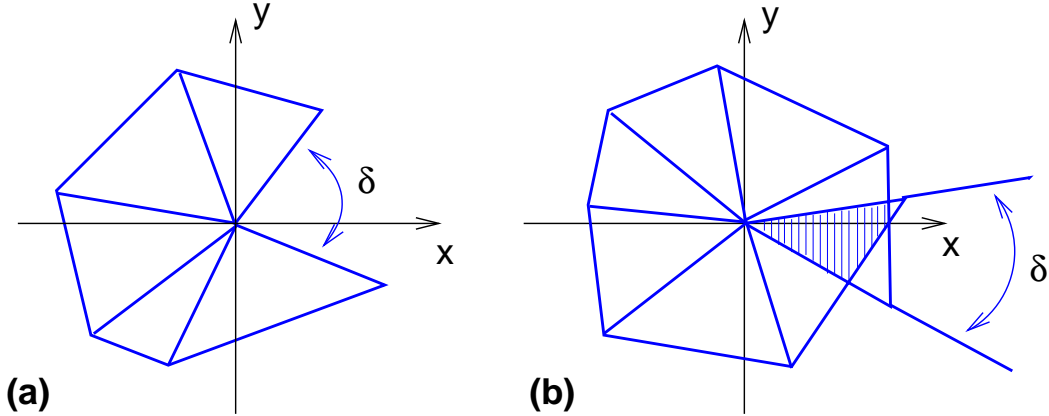}}}}
\caption[euangle]{Positive (a) and negative (b) 
Euclidean deficit angles $\delta$.}
\label{euangle}
\end{figure}

For the sake of completeness, we will describe some properties of
Lorentzian angles (or ``boosts'') which appear in the Regge 
scalar curvature as rotations about space-like bones (that is,
space-like links in 3d and space-like triangles in 4d). This summarizes
the treatment and conventions of \cite{sorkin}. The familiar form of
a contribution of a single bone $b$ to the total curvature and therefore
the action is
\begin{equation}
\Delta_{b} S={\rm volume}(b)\delta_{b},\;\;\;\;
\delta_{b} =2\pi -\sum_{i}\Theta_{bi},
\end{equation}
where the volume of the bone is by definition real and positive and
$\delta_{b}$ is the deficit angle around $b$. For the case of
Euclidean angles this is conveniently illustrated by a 
two-dimensional example, where the bone is simply a point with volume
1. A positive deficit angle $\delta$ (Fig.\ \ref{euangle}a) 
implies a positive Gaussian
curvature and therefore a positive contribution to the action,
whereas an ``excess'' angle $\delta$ contributes negatively
(Fig.\ \ref{euangle}b). 
\begin{figure}[t]
\centerline{\scalebox{0.6}{\rotatebox{0}{\includegraphics{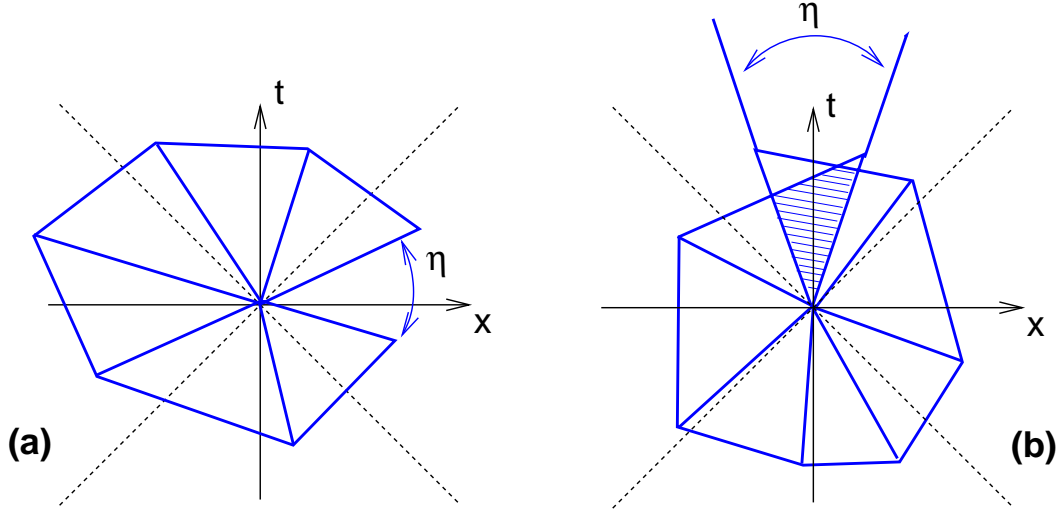}}}}
\caption[lorangle]{Positive space-like (a) and time-like (b) Lorentzian 
deficit angles $\eta$.}
\label{lorangle}
\end{figure}
Lorentzian angles are also defined to be additive, in such a way that 
a complete rotation around a space-like bone gives $2\pi$ in the flat 
case. However, note that the angles are in general {\it complex}, and 
can become arbitrarily large in the vicinity of the light cone. 
In their contribution to the action, we have to distinguish between two cases.
If the Lorentzian deficit angle $\delta\equiv\eta$ is space-like 
(Fig.\ \ref{lorangle}a), it contributes 
as $\Delta_{b} S={\rm volume}(b)\eta_{b}$, just
like in the Euclidean case. By contrast, if it is time-like 
(Fig.\ \ref{lorangle}b),
the deficit angle contributes with the opposite sign, that is,
as $\Delta_{b} S=-{\rm volume}(b)\eta_{b}$. Therefore, both a 
space-like defect and a time-like excess increase the action, where as 
a time-like defect or a space-like excess decrease it. 

The deficit angles in Sections 2.1 and 2.2 were calculated using
\begin{equation}
\cos\Theta =\frac{\langle \vec v_{1},\vec v_{2}\rangle }{
\langle \vec v_{1},\vec v_{1}\rangle^{\frac{1}{2}} 
\langle \vec v_{2},\vec v_{2}\rangle^{\frac{1}{2}} },\;\;\;\;
\sin\Theta =\frac{\sqrt{ 
\langle \vec v_{1},\vec v_{1}\rangle
\langle \vec v_{2},\vec v_{2}\rangle -
\langle \vec v_{1},\vec v_{2}\rangle^{2} }}{
\langle \vec v_{1},\vec v_{1}\rangle^{\frac{1}{2}} 
\langle \vec v_{2},\vec v_{2}\rangle^{\frac{1}{2}} },
\end{equation}
for pairs of vectors $\vec v_{1},\vec v_{2}$, and the flat
Minkowskian scalar product $\langle \cdot,\cdot \rangle$.
By definition, the square roots are positive imaginary for
negative arguments.

\section*{Appendix 2: Link-reflection positivity in 2d}

In this appendix we will demonstrate the link-reflection positivity of 
the discrete model of two-dimensional Lorentzian quantum gravity
introduced in \cite{al}. Recall that link reflection is the
reflection at a plane of half-integer $t$. We choose it to lie at
$t=1/2$ and fix the boundary spatial lengths at the initial time $t=-T+1$
and the final time $t=T$ to $l_{-T+1}=l_{T}=l$. In order to
describe a two-dimensional Lorentzian universe with these boundary 
conditions, we must not only specify the geometry of the spatial
sections (simply given by $l_{t}$, $-T+1\leq t\leq T$), but the
connectivity of the entire 2d triangulation. 

A convenient way of
parametrizing the connectivity that is symmetric with respect to
incoming and outgoing triangles at any given slice of constant $t$
is as follows. For any spatial slice at some integer time $t$, 
each of the $l_t$ spatial edges forms the base of one incoming
and one outgoing triangle. The geometry of the adjoining sandwiches
is determined by how these triangles are glued together
pairwise along their time-like edges. These gluing patterns 
correspond to distinct ordered partitions of the $l_t$ triangles 
(either above or below $t$) into $k$ groups, $1\leq k\leq l_t$. 
We denote the partitions collectively by $m(t)=\{ m_r(t),\ r=1,\ldots ,k\}$
for the incoming triangles and by 
$n(t)=\{ n_r(t),\ r=1,\ldots ,k'\}$ for the outgoing triangles. 
The constraints on these variables are obviously
$\sum_{r=1}^k m_r(t) =\sum_{r=1}^{k'} n_r(t) =l_t$ and a matching
condition for adjacent slices, namely, $k'(t)=k(t+1)$.\footnote{ 
This parametrization is closely related to the description of
2d Lorentzian gravity in terms of ``half-edges'' \cite{lottietal}.}
In terms of these variables, the (unmarked) one-step propagator
is given by
\begin{equation}
G_g(l_1,l_2)=g^{l_1+l_2} \sum_{k\geq 1}\; \frac{1}{k} \
\sum_{ {n_r,m_r\geq 1,\ r=1,\ldots,k \atop 
\sum_{q=1}^k n_q =l_1,\ \sum_{p=1}^k m_p =l_2 } } 1\;
=g^{l_1+l_2} \sum_{k\geq 1}\; \frac{1}{k} \
{l_1-1 \choose l_1-k } {l_2-1 \choose l_2-k},
\label{1step}
\end{equation}
where $g={\rm e}^{-\lambda}$ depends on the two-dimensional
cosmological constant $\lambda$. It is obvious from (\ref{1step}) that 
the propagator depends symmetrically on the local variables $m$ and $n$.
The partition function for the entire system is
\begin{equation}
Z_{2d}(g)=G_g(l_{-T+1},l_{-T+2}) \prod_{t=-T+2}^{T-1}\,
\sum_{l_t\geq 1} l_t\ G_{g}(l_t,l_{t+1}),
\end{equation}
in accordance with (\ref{iter}).

Following the discussion of link reflection in Sec.\ 6, we consider 
now functions $F(\{ m,n\} )$ that depend only on the geometry above 
the reflecting plane at $t=1/2$, i.e. on $m(t)$, $n(t)$, $t\geq 1$.
The reflection (\ref{lreflect}) acts on the partition data according 
to $\theta_l (m(t))=n(-t+1)$ and $\theta_l(n(t))=m(-t+1)$.
Without loss of generality, we can assume that $F$ depends
symmetrically on $m(t)$ and $n(t)$, $\forall t$. Computing the
expectation value (\ref{linkref}), we obtain
\begin{eqnarray}
&&\langle (\theta_l F)F\rangle\nonumber\\
&& \hspace{.3cm}= Z_{2d}^{-1} \sum_{l_0,l_1 \geq 1}
l_0 l_1 G_g(l_0,l_1)  \Bigl( G_g(l_{-T+1},l_{-T+2}) 
\prod_{t=-T+2}^{-1} \sum_{l_t\geq 1}
l_t G_g(l_t,l_{t+1}) (\theta_l F) \Bigr)\times\nonumber\\
&& \hspace{1.4cm} \Bigl( \prod_{t=1}^{T-2} \sum_{l_{t+1}\geq 1} 
G_g(l_t,l_{t+1}) l_{t+1}
G_g(l_{T-1},l_T) F \Bigr)\nonumber\\
&&\hspace{.3cm}=Z_{2d}^{-1} \sum_{l_0,l_1 \geq 1} l_0 l_1 \sum_{k\geq 1}\ 
\frac{1}{k}\ g^{l_0 +l_1}{l_0-1 \choose l_0-k}{l_1-1 \choose l_1-k }\ 
\overline{{\bf F}(l_0)} {\bf F}(l_1)\nonumber\\
&&\hspace{.3cm}=Z_{2d}^{-1} \sum_{k\geq 1}\ \frac{1}{k}\ 
\overline{{\cal F}(k)}{\cal F}(k).
\label{alongeq}
\end{eqnarray}
Since the last expression is a sum of positive terms, we have
hereby shown that 2d Lorentzian gravity is
link-reflection positive. 

\vfill\eject

\end{document}